\documentclass[12pt]{article}
\usepackage[a4paper,margin=1in]{geometry}
\usepackage{amsmath,amssymb,bm}
\usepackage{graphicx}
\usepackage{setspace}
\usepackage{hyperref}
\hypersetup{hidelinks}
\usepackage{tikz}

\title{Entropic Necks, Dynamic Crossovers, and Fragility in Supercooled Liquids}       
\author{\textbf{Biman Bagchi}\\
\small Solid State and Structural Chemistry Unit,\\
\small Indian Institute of Science, Bengaluru 560012, India\\
\small Email: profbiman@gmail.com; bbagchi@iisc.ac.in}

\date{}

\begin{document}
\maketitle

\begin{abstract}

The dramatic slowdown of dynamics in supercooled liquids is accompanied by a sequence of dynamical crossovers, most notably the transition from high-temperature collision-dominated transport to low-temperature activated structural relaxation. A particularly striking manifestation of this change is the crossover from Rosenfeld excess-entropy scaling to the Adam--Gibbs relation. In this work we develop a theoretical framework based on a configuration-space extension of Zwanzig's entropic-neck picture and combine it with a Mori--Zwanzig memory-function formalism to address anomalies of supercooled liquids.
The central idea is that structural relaxation is controlled by the narrowing of configurational pathways connecting metastable basins of the inherent-structure landscape. Starting from coupled slow variables describing intrabasin motion and neck fluctuations, we derive a reduced generalized Langevin description in which elimination of the neck coordinate generates a long-lived memory kernel and naturally leads to entropy-controlled activated dynamics. At high temperatures the neck is broad and readily accessible, yielding Rosenfeld-type transport governed primarily by local structural entropy. Upon cooling, progressive neck constriction produces an increasing entropy deficit, leading to Adam--Gibbs behavior and activated relaxation.
Within this picture, fragility acquires a simple geometric interpretation: fragile liquids are characterized by a rapid collapse of the effective configurational neck with decreasing temperature, whereas strong liquids exhibit a much slower evolution of accessible pathways. The theory further provides a microscopic interpretation of the crossover temperature, clarifies its relation to mode-coupling ideas, and naturally extends to nonequilibrium cooling where the same neck-controlled dynamics governs fictive-temperature evolution. The entropic-neck framework thus offers a physically transparent route for understanding crossover phenomena, activated relaxation, and fragility in supercooled liquids. The framework does not by itself compute the configurational entropy, mismatch penalty, or cooperative length from microscopic interactions; in this respect, RFOT provides a more microscopic thermodynamic route, while the present approach supplies a complementary dynamical and geometrical interpretation.

\end{abstract}

\newpage

\section{Introduction}
Understanding the origin of the spectacular growth of relaxation times and viscosities in supercooled liquids—typically by 14--15 orders of magnitude over modest temperature intervals—remains  one of the central challenges in condensed-matter physics
\cite{Kauzmann1948,Angell1997review, GibbsDiMarzio1958, AdamGibbs1965,
Angell1995,Angell2000, Ediger2000, EdigerAngellNagel1996,Nagel1998,CRR_KobAndersen2023}. Because pair structure, quantified by $g(r)$ or $S(q)$, changes only modestly upon supercooling, it has long been suspected that the dominant cause of slowdown is a progressive restriction of the liquid’s exploration of configuration space. 
It is therefore natural that two of the most influential approaches relate \emph{entropy} to \emph{dynamics}.

At high or moderately supercooled temperatures, Rosenfeld’s excess-entropy scaling (RS) links reduced transport coefficients to the excess entropy $S_{\rm ex}$, often well approximated by the pair contribution $S_2[g(r)]$ \cite{Rosenfeld1977,Chakravarty2006,Chakravarty2007,Chakravarty2011}.
At low temperatures, the Adam–Gibbs (AG) framework connects relaxation to the configurational entropy $S_c(T)$ and the growth of cooperatively rearranging regions (CRRs). [4,5]
Despite the wide empirical support for both relationships, a coherent mechanistic explanation of the \emph{crossover} between these regimes remains incomplete.

In this paper, we develop a unified theoretical framework for structural
relaxation in supercooled liquids based on Zwanzig’s entropic bottleneck
picture, viewed as diffusion through a channel of varying cross section
in configuration space. \cite{Zwanzig1992,ZhouZwanzig1991,Zwanzig2001,Bagchi2012,MachtaZwanzigPRL}
In Zwanzig's model, developed on the inherent structure picture, the system explores the potential energy surface spanned by the configurations of the system.
We argue that this geometric description provides
a natural organizing principle for understanding the crossover from
high-temperature, entropy-controlled transport to low-temperature,
activated dynamics. Within this framework, several seemingly distinct
approaches can be viewed as different manifestations of a common underlying
constraint: the progressive narrowing and reduced persistence of accessible
phase-space pathways.

Building on this picture, we obtain several key results. First, we show that
the Adam--Gibbs relation emerges naturally from an entropy-deficit description
of viable configurational pathways, providing a direct geometric interpretation
of activated relaxation. Second, we formulate the problem using the
Mori--Zwanzig projection operator technique with two slow variables: a
progress coordinate describing motion between basins and a neck variable
capturing fluctuations in configurational connectivity. Elimination of the
neck variable leads to a closed generalized Langevin equation in which the
slow memory kernel is governed by the dynamics of the entropic bottleneck,
thereby embedding Adam--Gibbs behavior within a microscopic dynamical
framework.

Third, we introduce a simple but physically transparent decomposition of the
relaxation rate as the product of the number of available inter-basin
connections and the probability that a given connection is dynamically
viable. This factorization provides a natural explanation for the crossover
from Rosenfeld excess-entropy scaling at high temperatures to Adam--Gibbs
behavior at low temperatures. Finally, we relate the progressive collapse
of viable neck configurations to the fragility of glass-forming liquids,
thereby connecting geometric constraints in configuration space to the
observed diversity of dynamical behavior.

Taken together, these results suggest that the central bottleneck in glassy
dynamics is entropic in origin and arises from the scarcity and temporal
persistence of viable pathways in configuration space. The present formulation
thus provides a unified perspective linking Zwanzig’s geometric picture,
entropy-controlled transport relations, and modern theories of activated
relaxation.

Before developing this viewpoint in Sec.~2, we briefly review the main entropy-based 
approaches and dynamical theories relevant to the problem.

\subsection{Configurational entropy, Kauzmann paradox, and the Adam--Gibbs relation}

Kauzmann’s classic analysis \cite{Kauzmann1948} revealed that the entropy of a supercooled liquid, if extrapolated below the glass transition temperature $T_g$, decreases more rapidly than the entropy of the corresponding crystal.  
This implies a hypothetical temperature $T_K$ at which the supercooled liquid’s entropy would equal (or fall below) the crystal entropy—an apparent ``entropy crisis'' \cite{Angell1997review}.  
Although kinetic arrest or crystallization prevents experimental access to $T_K$, the idea that configurational entropy controls deep supercooling remains central.

Gibbs and DiMarzio \cite{GibbsDiMarzio1958} gave the first thermodynamic formulation, interpreting $S_c$ as the logarithm of the number of amorphous basins accessible at a given energy.  
Adam and Gibbs \cite{AdamGibbs1965} provided the dynamical extension: relaxation occurs through cooperative rearrangements of regions of size $z^\star$, and the activation free energy is proportional to $z^\star$.  
This yields the celebrated relation

\begin{equation}
\tau(T)
= \tau_0 \exp\!\left[\frac{C}{T\,S_c(T)}\right],
\label{eq:AG}
\end{equation}

which holds well for many fragile liquids.

Modern simulations have emphasized both the usefulness and the limitations of
inherent-structure-based or vibrational--configurational estimates of \(S_c(T)\).
Such estimates often capture the qualitative decrease of configurational entropy
upon supercooling, but they need not coincide quantitatively with more rigorous
definitions used in the RFOT context. Point-to-set correlations, finite-size scaling,
and four-point structure factors further support the growth of cooperative length
scales.  \cite{BerthierPNAS2017, CRR_KobAndersen2023,BerthierPNAS2017}.  
Nevertheless, quantitative implementation of AG remains nontrivial: $S_c(T)$ depends on basin definitions, anharmonic contributions, and sampling procedures; different dynamical observables may decouple; and the prefactor $\tau_0$ can be mildly temperature dependent.

\subsection{Excess entropy, pair structure, and Rosenfeld scaling}

At higher temperatures, and in liquids whose basins are broad and quasi-harmonic, dynamical slowdown is much more modest and controlled by \emph{local} geometric constraints.  
In this regime, Rosenfeld discovered that reduced transport coefficients collapse onto universal curves when plotted against the excess entropy $S_{\rm ex}=S-S_{\rm id}$ \cite{Rosenfeld1977}.  
For diffusion,

\begin{equation}
D^\ast = A \exp(\alpha S_{\rm ex}),
\label{eq:RS}
\end{equation}

where $D^\ast$ is a dimensionless transport coefficient and $A$, $\alpha$ depend weakly on the interaction potential.

The dominant contribution to $S_{\rm ex}$ in simple liquids comes from the pair-correlation entropy,

\begin{equation}
S_2
= -2\pi\rho\!\int_0^\infty 
\left[g(r)\ln g(r) - (g(r)-1)\right] r^2 dr,
\label{eq:S2}
\end{equation}

which depends only on $g(r)$.  
This makes RS particularly appealing for simulations: it establishes a direct quantitative bridge between local structure and transport.

Chakravarty and co-workers extensively explored this bridge in a series of papers from 2002–2014.  
They demonstrated that many thermodynamic and dynamic anomalies of water-like liquids can be mapped onto contours of constant $S_2$ or $S_{\rm ex}$, and that Rosenfeld scaling often holds even in systems with directional bonding provided structural contributions are decomposed appropriately.  
Deviations from RS correlate strongly with changes in the shape of $g(r)$ and the development of intermediate-range order.
At high or moderately supercooled temperatures, Rosenfeld’s excess-entropy scaling (RS) links reduced transport coefficients to the excess entropy $S_{\rm ex}$, often well approximated by the pair contribution $S_2[g(r)]$.
\cite{Rosenfeld1977,Chakravarty2006,Chakravarty2007,Chakravarty2011}

Bhattacharyya and collaborators developed a complementary perspective: by studying how $g(r)$ evolves with cooling across different model liquids, they showed that the breakdown of RS correlates with specific structural changes, such as sharpening of the first peak or splitting of the second peak. \cite{BanerjeeBhattacharyya2016, NandiBhattacharyya2017}
These effects reflect the emergence of more rugged and anharmonic energy basins.  
In fragile liquids RS tends to fail earlier because local structural measures cannot account for the increasingly collective nature of relaxation, whereas in strong liquids RS often persists deeper into supercooling.

 At high temperature the dynamics are dominated by local intrabasin motion: collisional dynamics, 
 cage exploration, and short-range structural rearrangements.
 These observations suggest that Rosenfeld scaling is most appropriate in the
high-temperature or moderately supercooled regime, where transport is still
controlled primarily by local structural correlations. In this regime, changes in
\(g(r)\), and hence in the pair entropy \(S_2\), provide a useful measure of the
local packing constraints that influence diffusion and viscosity. This local
Rosenfeld regime should be distinguished from the deeply supercooled regime,
where relaxation is increasingly controlled by rare interbasin passages and by
the configurational entropy associated with the number of available amorphous
basins.

\subsection{Dynamical theories of supercooled liquids: MCT, RFOT, and the BBW synthesis}

The structural observations discussed above raise a broader question: how do local correlations ultimately generate the dramatic dynamical slowdown of supercooled liquids?  
Several theoretical frameworks have been developed to address this problem.

\emph{Mode-coupling theory} (MCT) describes the early stages of dynamical slowdown in terms of nonlinear feedback between density fluctuations \cite{Goetze_book}.  

In this approach, particles become transiently trapped in cages formed by their neighbors, leading to a dynamical crossover at a temperature $T_c$ above the laboratory glass transition temperature $T_g$.  
MCT successfully captures many features of dynamics in the moderately supercooled regime but does not account for activated processes that dominate at lower temperatures.

A complementary perspective is provided by the \emph{Random First-Order
Transition} (RFOT) theory, which emerged from a sequence of developments
combining ideas from mean-field spin-glass theory, structural glasses,
and cooperative dynamics
\cite{KirkpatrickThirumalaiWolynes1987,KirkpatrickThirumalaiWolynes1989,MezardParisi1999,BouchaudBiroli2004,LubchenkoWolynes2007}.
RFOT interprets glassy dynamics in terms of a multiplicity of metastable
amorphous states separated by free-energy barriers, with relaxation
occurring through activated rearrangements between these states. In this
framework, the competition between interfacial mismatch and the
configurational entropy of alternative states leads to a characteristic
cooperative length scale and an associated activation free energy that
grows upon supercooling. This picture provides a thermodynamic route to
the Adam--Gibbs relation and has been further developed and refined in a
number of subsequent works.

In the present context, these activated rearrangements correspond to
the rare configurational bottlenecks (entropic necks) that control the
slow component of the memory kernel.

 Bhattacharyya, Bagchi and Wolynes proposed an important synthesis of these viewpoints by combining MCT-like collisional dynamics with RFOT-like activated hopping processes \cite{BBW_JCP_2004,BBW_PRE_2005,BBW_PNAS_2008}.  
In this picture, the dynamics of supercooled liquids involve a crossover from collision-dominated motion described by MCT to activated interbasin transitions governed by configurational entropy.BBW assumed that the memory kernel of MCT, the wave number and frequency dependent rate of relaxation of the dynamic structure factor can be described as a sum of the usual MCT term and a RFOT hopping term. The former is determined self-consistently the non-linear MCT equation. 

Despite this progress, the precise relationship between Rosenfeld scaling, MCT, and RFOT remains unclear.  
In particular, it is not obvious whether Rosenfeld scaling should be viewed as a high-temperature structural limit of these theories or as a manifestation of more general geometric constraints on the accessible configuration-space volume.

In the following sections we show that Zwanzig’s entropic bottleneck picture provides a natural framework for understanding these connections.

\subsection{Crossover, fragility, and a geometry-based entropic-neck picture}

At high temperatures, transport correlates well with excess entropy, and Rosenfeld scaling holds.  
Upon deeper supercooling—especially in fragile liquids—the dynamics become heterogeneous and activated, and relaxation times correlate more naturally with the configurational entropy and the AG relation.  
The temperature or density at which RS breaks down, MCT begins to fail, and AG becomes appropriate is not universal; it correlates strongly with the liquid’s fragility.

Fragile liquids typically exhibit an earlier breakdown of Rosenfeld scaling because the effective configuration-space bottle-neck narrows rapidly with decreasing temperature, forcing increasingly collective rearrangements. In contrast, strong liquids display a slower collapse of the neck entropy and therefore retain Rosenfeld-type correlations between transport and excess entropy over a wider temperature range.
The Kob–Andersen mixture displays both regimes depending on temperature, making it a prototypical crossover system.\cite{KobAndersen1995}

Despite the empirical success of RS and AG, and the partial theoretical synthesis offered by BBW and RFOT, a fundamental mechanistic question remains unresolved:
\emph{How can the same microscopic system transition from a regime governed by local, two-body geometric entropy to a regime governed by collective, configurational entropy?}

In this work we propose that Zwanzig’s entropic-bottleneck picture offers an economical and physically transparent understanding of the crossover.  
In Sec.~2 we reinterpret diffusion in configuration space as motion through an effective ``channel'' whose cross section encodes the number of locally accessible microstates.  
At high temperature, the relevant bottleneck involves only local degrees of freedom and naturally 
yields Rosenfeld-like scaling.  
At low temperature, the effective cross section is controlled by the simultaneous rearrangement of 
many degrees of freedom, and the bottleneck becomes a \emph{collective entropic neck} whose width is 
controlled by $S_c(T)$, leading to AG-type behavior.  
This geometric framework provides a unified language for understanding how RS, MCT, AG, fragility, 
and RFOT emerge as limiting descriptions of the same underlying mechanism.

A central result of the present work is the formulation of a coupled
Mori--Zwanzig description in terms of two slow variables: a progress
coordinate \(X\), describing motion between metastable basins, and a neck
variable \(Q\), characterizing the instantaneous configurational
connectivity. While the geometric picture suggests that relaxation is
controlled by rare bottleneck configurations, the Mori--Zwanzig framework
provides the corresponding dynamical realization. In particular, the
elimination of the neck dynamical variable \(Q\) generates a non-Markovian memory
kernel for \(X\), whose long-time behavior is governed by the statistics
and persistence of entropic bottlenecks. This leads directly to a
relaxation timescale controlled by an entropy deficit, yielding the
characteristic \(\tau\)–\(s_c\) dependence of the Adam--Gibbs relation.
Thus, the emergence of activated dynamics is not imposed phenomenologically
but arises naturally from the dynamics of fluctuating configurational
connectivity.

The remainder of the paper is organized as follows. In Section~2 we
introduce Zwanzig's entropic-neck picture and show how diffusion through
a configuration-space constriction can be described using the
Fick--Jacobs reduction, leading to an entropic barrier associated with
the narrowing of accessible phase-space volume. Section~3 we discuss the
Mori-Zwanzig formulation of the dynamics of the present problem.
In Section~4 we discuss how the Adam--Gibbs relation
emerges naturally from the entropy deficit associated with a narrowing
neck and how Rosenfeld scaling appears as the high-temperature limit
where the neck is effectively open. Section~5 develops a parallel
memory-function decomposition in which intrabasin collisional dynamics
and interbasin neck-controlled dynamics contribute independently to the
generalized friction. Section~6 compares the entropic-neck mechanism
with the nucleation-based picture of the Random First-Order Transition
(RFOT) theory and clarifies how the two frameworks compute complementary
probabilities associated with structural relaxation. Finally,
Section~7 connects with other theories. In S1, we discuss how to relate 
the fictive-temperature dynamics to an enthalpy
memory kernel and expresses this kernel in terms of density correlation
functions using a mode-coupling representation, thereby relating the
nonequilibrium breakaway temperature to the mode-coupling crossover.

\section{Zwanzig’s Entropic Neck Picture}
  Zwanzig introduced several simple, often analytically tractable models (i) to understand the emergence of diffusion in a deterministic system, such as motion of a point particle in a regular Lorentz gas of fixed scatterers and (ii) to model motion of a system in its energy landscape, motivated by the inherent structure analysis of supercooled liquids. \cite{Zwanzig1992,Zwanzig2001,Bagchi2012,MachtaZwanzigPRL}
  
  In the first model, the entropic trap induces memory loss of the trapped particle, leading to a random walk in the configuration space. Physics is far more complex in the second model where the neck needs to appear in the configuration space. Thus, we need to consider the fluctuating neck.

  This reminds us of a model well-known in protein dynamics. Agmon and Hopfield  introduced such a model to explain non-exponentiality observed in enzyme kinetics.\cite{AgmonHopfield1983} The neck was assumed to follow a damped harmonic motion, and escape happened when the opening of the neck was larger than certain critical value. This in turn can be related to a model by Eshelby where elastic energy was introduced to describe creation of strain in solids.\cite{Eshelby}
  
  In this entropic neck picture, the particle (the system) needs to pass through a constriction or a narrow region to continue diffusive motion. In the short time, the particle may execute Brownian motion in a region characterized by substantial degrees of freedom, but the region is separated from other regions by a constriction, or a neck, that the particle needs to pass through, as mentioned above.

The narrow constriction has much lower degrees of freedom, and hence lower entropy than the broad regions on all sides.
But the neck width can fluctuate, as discussed later.

Zwanzig used an earlier analysis by Fick and Jacobs to show that the existence of the neck gives rise to a large entropic barrier, and this allows us to model diffusion as a random walk where passing through the neck is modeled as an activated process.\cite{FickJacobs1967}

This simple model allows a conceptual visualization where the appearance of the neck provides a change in the diffusion mechanism.
We discuss below how a memory function formalism can be employed to obtain a unified description of diffusion in such systems.

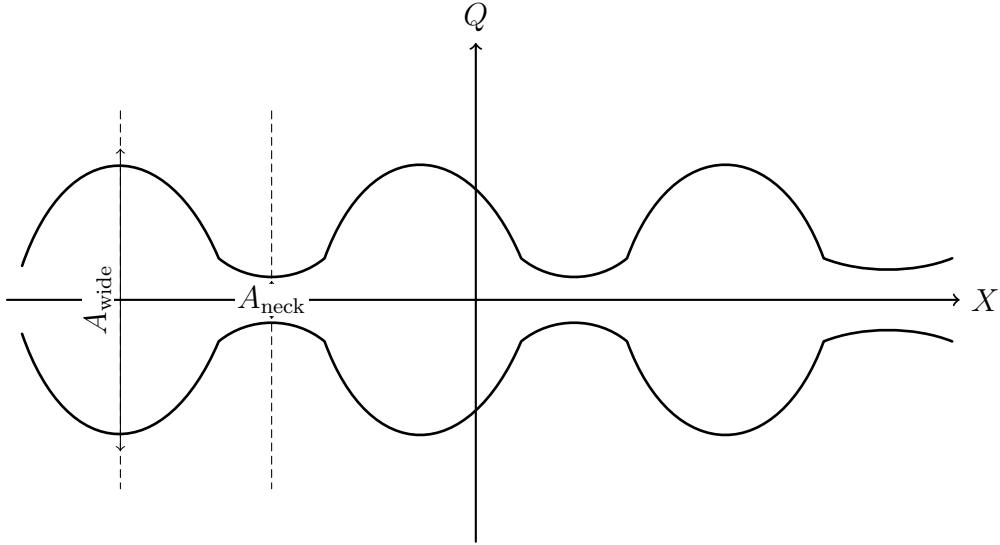
\begin{figure}[htbp]
\centering
\begin{tikzpicture}[scale=1.0, line cap=round, line join=round]

\draw[->, thick] (-6.2,0) -- (6.4,0) node[anchor=west] {$X$};
\draw[->, thick] (0,-3.2) -- (0,3.4) node[anchor=south] {$Q$};

\draw[line width=1]
(-6.0,0.45)
 .. controls (-5.4,2.2) and (-4.1,2.2) .. (-3.4,0.55)
 .. controls (-3.0,0.22) and (-2.4,0.22) .. (-2.0,0.55)
 .. controls (-1.4,2.2) and (-0.1,2.2) .. (0.6,0.55)
 .. controls (1.0,0.22) and (1.6,0.22) .. (2.0,0.55)
 .. controls (2.6,2.2) and (4.0,2.2) .. (4.6,0.55)
 .. controls (5.1,0.35) and (5.8,0.35) .. (6.3,0.55);

\draw[line width=1]
(-6.0,-0.45)
 .. controls (-5.4,-2.2) and (-4.1,-2.2) .. (-3.4,-0.55)
 .. controls (-3.0,-0.22) and (-2.4,-0.22) .. (-2.0,-0.55)
 .. controls (-1.4,-2.2) and (-0.1,-2.2) .. (0.6,-0.55)
 .. controls (1.0,-0.22) and (1.6,-0.22) .. (2.0,-0.55)
 .. controls (2.6,-2.2) and (4.0,-2.2) .. (4.6,-0.55)
 .. controls (5.1,-0.35) and (5.8,-0.35) .. (6.3,-0.55);

\draw[densely dashed] (-2.7,2.5) -- (-2.7,-2.5);
\draw[<->] (-2.7,0.25) -- (-2.7,-0.25);
\node[fill=white,inner sep=1pt] at (-2.7,0) {$A_{\rm neck}$};

\draw[densely dashed] (-4.7,2.5) -- (-4.7,-2.5);
\draw[<->] (-4.7,2.0) -- (-4.7,-2.0);
\node[fill=white,inner sep=1pt,rotate=90] at (-5.0,0) {$A_{\rm wide}$};

\end{tikzpicture}


\caption{Schematic representation of Zwanzig's entropic bottleneck picture in configuration space. 
The coordinate $X$ denotes progress between neighboring metastable basins, while $Q$ denotes a schematic transverse neck coordinate that represents the instantaneous openness or configurational accessibility of the bottleneck. 
Wide regions correspond to basins and narrow but finite regions correspond to entropic necks. 
The local width $A(X)$ measures the effective number of accessible configurations at a given $X$ and defines an entropic potential $F(X)=-k_{\rm B}T\ln A(X)$. 
Thus constricted regions act as entropic barriers that reduce the probability of interbasin passage.}

\label{fig:entropic_neck}

\end{figure}
%
%
%
\subsection{Fick--Jacobs Reduction and the Origin of the Entropic Potential}

The effect of the constriction sketched in Figure 1
can be made quantitative using the Fick--Jacobs reduction for diffusion
in a narrow tube of variable cross section $A(x)$.\cite{FickJacobs1967}

For an overdamped Brownian particle with diffusion coefficient $D_0$,
the two-dimensional Smoluchowski equation
\[
\frac{\partial P}{\partial t}
= D_0\nabla^2 P
\]
reduces, after integrating over the fast transverse coordinate, to
\begin{equation}
\frac{\partial p(x,t)}{\partial t}
 = D_0\,\frac{\partial}{\partial x}
   \!\left[
     A(x)\frac{\partial}{\partial x}
     \!\left(\frac{p(x,t)}{A(x)}\right)
   \right],
\label{eq:FJ}
\end{equation}
where $p(x,t)=\int_{A(x)} P(x,y,t)\,dy$ is the marginal probability.\cite{FickJacobs1967}

Equation~\eqref{eq:FJ} can be rewritten as a one--dimensional
Smoluchowski equation with an \emph{effective free energy}
\begin{equation}
F(x) = -k_B T \ln A(x),
\label{eq:F_entropic}
\end{equation}
so that
\begin{equation}
\frac{\partial p}{\partial t}
= D_0\,\frac{\partial}{\partial x}
  \left[
   e^{-\beta F(x)}\frac{\partial}{\partial x}
   \!\left(e^{\beta F(x)}p\right)
  \right].
\end{equation}
Hence, a narrow region (small $A$) acts as an entropic barrier of height
\begin{equation}
\Delta F_{\text{neck}} = k_B T
   \ln\!\left(\frac{A_{\text{wide}}}{A_{\text{neck}}}\right).
\end{equation}

This expression provides the microscopic basis for Zwanzig’s picture:
regions of small neck available cross section correspond to higher free energy,
purely through a loss of entropy, and therefore limit diffusive transport
between adjacent wide basins.
%

\subsection{Random walk through the constriction}
\label{subsec:random_walk_constriction}

Zwanzig analyzed transport in a tube of varying cross section $A(x)$ by using the earlier Fick--Jacobs reduction. In this description the transverse degrees of freedom are locally equilibrated, and motion along the longitudinal coordinate $x$ occurs in an effective entropic potential. The corresponding free energy is
\begin{equation}
F(x)
=
-k_{\rm B}T\ln A(x).
\label{eq:Fick_Jacobs_entropic_potential}
\end{equation}
Thus a narrow region of the tube has a smaller number of transverse microstates and therefore a higher effective free energy. If a wide region has cross section $A_{\rm wide}$ and the constriction has cross section $A_{\rm neck}$, the entropic free-energy barrier is
\begin{equation}
\Delta F_{\rm neck}
=
k_{\rm B}T
\ln
\left(
\frac{A_{\rm wide}}
{A_{\rm neck}}
\right).
\label{eq:entropic_barrier_area}
\end{equation}
The probability of crossing the neck is therefore reduced by the Boltzmann factor
\begin{equation}
p_{\rm neck}
=
\exp
\left[
-\beta \Delta F_{\rm neck}
\right],
\label{eq:p_neck_area}
\end{equation}
where $\beta=(k_{\rm B}T)^{-1}$. This is the central physical content of the Zwanzig--Fick--Jacobs construction: a purely entropic reduction of available cross section appears as an effective free-energy barrier.

We now generalize this picture from an ordinary geometrical tube to the configuration space of a supercooled liquid. In a glass-forming liquid, the system does not move through a literal tube. Rather, it explores a high-dimensional landscape of metastable basins connected by rare, narrow regions of configurational connectivity. These narrow regions play the role of entropic necks. The transverse area $A(x)$ in the tube problem is therefore replaced by a configurational measure.

Let $W_{\rm basin}(T)$ denote the number, or measure, of intrabasin configurations accessible to a cooperatively rearranging region. Let $W_{\rm neck}(T)$ denote the much smaller measure of configurations that form viable ``doorway'' states leading from one basin to another. The scarcity of such doorway configurations defines the dimensionless entropic-neck deficit
\begin{equation}
\Delta\Sigma_{\rm neck}(T)
=
\ln
\left[
\frac{W_{\rm basin}(T)}
{W_{\rm neck}(T)}
\right].
\label{eq:DeltaSigma_neck_definition}
\end{equation}
The corresponding entropy loss is
\begin{equation}
\Delta S_{\rm neck}(T)
=
k_{\rm B}\Delta\Sigma_{\rm neck}(T).
\label{eq:DeltaS_neck_definition}
\end{equation}
Equivalently, the entropic part of the barrier is
\begin{equation}
\Delta F_{\rm neck}(T)
=
T\Delta S_{\rm neck}(T)
=
k_{\rm B}T\Delta\Sigma_{\rm neck}(T).
\label{eq:DeltaF_neck_definition}
\end{equation}
Consequently, the dimensionless activation factor associated with finding and traversing the neck is
\begin{equation}
\frac{\Delta F_{\rm neck}(T)}
{k_{\rm B}T}
=
\Delta\Sigma_{\rm neck}(T)
=
\ln
\left[
\frac{W_{\rm basin}(T)}
{W_{\rm neck}(T)}
\right].
\label{eq:dimensionless_neck_barrier}
\end{equation}

If one keeps the original geometrical language of Zwanzig's tube model, $W_{\rm basin}$ is analogous to the cross section of the wide part of the tube, while $W_{\rm neck}$ is analogous to the cross section of the constriction. In that limit Eq.~(\ref{eq:DeltaSigma_neck_definition}) reduces to
\begin{equation}
\Delta\Sigma_{\rm neck}
\simeq
\ln
\left(
\frac{A_{\rm wide}}
{A_{\rm neck}}
\right).
\label{eq:DeltaSigma_area_limit}
\end{equation}
For glassy dynamics, however, the important point is that $W_{\rm neck}$ is not merely a static geometrical area. It represents the measure of configurations that are both sufficiently open and sufficiently persistent to allow interbasin escape. This observation motivates the introduction, in the following sections, of a fluctuating neck variable whose dynamics generates an additional slow memory kernel.
%
\subsection{Who participates in forming the neck? Participation ratio }
 The first thing to note is that not  all $z$ degrees of freedom participate in creating the neck . We further
   quantify the neck as a product of two factors.
\begin{equation}
n_{\rm act}(T)\;=\;\phi(T)\,z^\star(T),\qquad 0<\phi\le 1,
\end{equation}
Here $n_{act}(T)$  be  the \emph{ minimum number of active coordinates} that must rearrange coherently to realize a viable neck, where $z^\star= k_{B} \Delta\Sigma /s_c$ is the minimal cooperative size from connectivity (Eq.~\eqref{eq:zstar}).  This $z^{*}$ has the same meaning as in Adam-Gibbs formulation.[3-5]

The factor $\phi(T)$ is a \emph{participation ratio}: it captures localization of the gate (e.g., boundary cluster or string-like motion) and can be smaller than unity. $\phi(T)$ can also contain an enthalpic component involved in the gate opening.

A clear picture of this picture can be obtained within the Agmon-Hopfield model where the lid or the gate in an enzyme must open for the substrate to go in, and again the gate must open for the substrate to come out. There are only a certain number $n^{*}$  of  enzyme states that participate in this process; $n^{*}$ can be much smaller than the total number of states of the enzyme.\cite{AgmonHopfield1983}

In the case of reactions like enzyme kinetics, the activation free energy may be decomposed into a local ``gate-creation'' cost plus any elastic/mismatch cost that propagates beyond the gate:
\begin{equation}
\Delta F^\ddagger
\;=\;\varepsilon_{\rm loc}\,n_{\rm actlocal}
\;+\;{E_{\rm el}(z^\star)}_{\text{elastic mismatch cost}} .
\label{eq:barrier_decompose}
\end{equation}
%


Here $n_{\rm actlocal}$ denotes the number of degrees of freedom that 
participate locally in forming the ``gate'' (the narrow entropic neck).
This is a localized cluster (typically much smaller than the full AG 
cooperative size $z^\star$ consisting of soft boundary sites or 
string-like coordinates whose rearrangement creates a viable escape 
direction. In contrast, the elastic accommodation of this local 
rearrangement extends into the surrounding region of size $z^\star$, 
giving rise to the mismatch term $E_{\rm el}(z^\star)$.

\subsection {Agmon-Hopfield Model of protein dynamics}

We now discuss a potentially interesting example with relevance to Agmon-Hopfield type of models. We call it Eshelby-assisted gate (elastic field extends into the core).

If forming the gate imposes a shear-like strain that must be elastically accommodated by a volume of order $V^\star\propto z^\star$, the mismatch cost has the Eshelby form \cite{Eshelby}
\begin{equation}
E_{\rm el}(z^\star)\;\simeq\;\frac{1}{2}G_\infty\,\gamma^2 V^\star
\;\propto\; z^\star,
\end{equation}
with $G_\infty$ the high-frequency shear modulus and $\gamma$ a (small) transformation strain. Then $\Delta F^\ddagger \propto z^\star$ regardless of locality of the gate, yielding
\begin{equation}
\ln\tau \;\sim\; \frac{\text{const}}{k_B T}\,\frac{z^\star}{1}
\;=\; \frac{\text{const}}{k_B T}\,\frac{\Delta\Sigma}{s_c},
\end{equation}
again the AG dependence. Physically, even if the \emph{gate} is local, the elastic accommodation is volumetric, restoring the linear-in-$z^\star$ scaling.
While the Agmon-Hopfield model of opening of gate is illustrative, opening of such a gate in the glassy liquid obviously not feasible. One can imagine the diffusing particle as the system moving in a quenched configuration (like the basins that constitute minima of inherent structures) --- the system moves from one metastable minima to another. The gate in such a description is the energy maximum or a barrier in the multidimensional configuration space, and a large number of atoms/molecules move something close to Lindemann length. In this picture, self-diffusion occurs through structural relaxation that accompanies the transition from one metastable minimum to the other.
In the RFOT picture, the crossing of barrier occurs through nucleation. The barrier appears because of the competition between entropic gain by transforming presumably to a higher entropic state but resisted by the mismatch at the surface which is accounted for by a surface tension. One obtain the surface term from a spin glass expression for surface mismatch between two metastable  (the initial and the final) states. The barrier is obtained by following a standard procedure. At the end one finds a barrier that is inversely proportional to the configuration entropy of the final state (or, the entropy difference between the initial and the final state).


\section{Coupled Mori--Zwanzig equations for the progress coordinate and the neck variable}

The preceding discussion has established that structural relaxation in
supercooled liquids is governed by the formation and persistence of
entropic bottlenecks connecting metastable basins. While this picture
provides a transparent geometric and thermodynamic interpretation of
activated dynamics, it does not by itself yield a dynamical equation of
motion. In particular, the slowing down must ultimately be understood
in terms of memory effects arising from the coupling of slow collective
variables to the remaining microscopic degrees of freedom.

To develop such a description, it is natural to employ the
Mori--Zwanzig projection operator formalism, which provides a systematic
route for deriving exact equations of motion for a set of chosen slow
variables by eliminating all other degrees of freedom
\cite{Zwanzig_book2001,Bagchi_book2023}. 

In the present context, the essential
physics involves two coupled slow variables: a progress coordinate \(X\),
which describes motion between neighboring basins in configuration space,
and a neck variable \(Q\), which characterizes the instantaneous
connectivity or openness of the configurational bottleneck. The inclusion
of both variables is crucial, as the motion along \(X\) is dynamically
constrained by the fluctuating availability of viable pathways encoded in
\(Q\).
$X(\Gamma)$ may be viewed as a progress coordinate, or reaction coordinate, that distinguishes neighboring metastable basins. Examples include a committor-like coordinate, an overlap with a reference inherent structure, or a low-dimensional collective coordinate constructed from density or displacement fields. Similarly, $Q(\Gamma)$ is not the volume of the neck. Rather, it is a scalar collective variable that measures the instantaneous openness or viability of the bottleneck. It may be represented schematically as a smooth functional of the microscopic density field, for example as an overlap with a constriction or rearrangement mode.
Thus the projection formalism is not being applied directly to phase-space volumes. It is applied to collective variables whose fluctuations encode the accessibility of the bottleneck.

 A microscopic realization of \(Q\) may be constructed from slow density or overlap
fields, providing a formal route by which the neck variable can be related to the
same slow structural modes that appear in mode-coupling descriptions.

The Mori--Zwanzig formalism allows us to derive a coupled set of
generalized Langevin equations for these variables, in which the influence
of the eliminated microscopic degrees of freedom appears through
memory kernels and fluctuating forces.\cite{Zwanzig2001, Zwanzig_book2001, Bagchi_book2023} \textit{As will be shown below, the
elimination of the neck variable \(Q\) generates a slow, non-Markovian
contribution to the memory kernel for \(X\), providing a direct dynamical
mechanism for the emergence of activated relaxation and its connection to
configurational entropy.}

We consider the full microscopic phase-space point
\begin{equation}
\Gamma = \{\mathbf{r}^N,\mathbf{p}^N\}
\end{equation}
evolving under the Liouville operator \(i{\cal L}\), so that for any dynamical
variable \(A(\Gamma)\),
\begin{equation}
\frac{d}{dt}A(t)= i{\cal L}A(t),
\qquad
A(t)= e^{i{\cal L}t}A(0).
\label{eq:Liouville_evolution}
\end{equation}

Our goal is to derive an effective description for two slow collective variables:

\begin{enumerate}
\item the longitudinal progress coordinate \(X(\Gamma)\), which parametrizes motion
between neighboring metastable basins in configuration space, and
\item a neck or gate variable \(Q(\Gamma)\), which characterizes the instantaneous
accessibility of the narrow configurational bottleneck connecting these basins.
\end{enumerate}

For the inertial Mori formulation, the natural relevant set is taken to be
\begin{equation}
\mathbf{A}(\Gamma)=
\begin{pmatrix}
X(\Gamma)\\[3pt]
P(\Gamma)\\[3pt]
Q(\Gamma)
\end{pmatrix},
\label{eq:relevant_vector}
\end{equation}
where \(P\) is the momentum conjugate to \(X\), defined so that
\begin{equation}
\dot X = \frac{P}{M},
\label{eq:Xdot_def}
\end{equation}
with \(M\) an effective mass associated with the collective coordinate \(X\).

We define the Mori projection operator \({\cal P}\) acting on any phase-space variable
\(B(\Gamma)\) by
\begin{equation}
{\cal P}B
=
\sum_{\alpha,\beta}
(B,A_\alpha)\,
\big[(\mathbf{A},\mathbf{A})^{-1}\big]_{\alpha\beta}\,
A_\beta,
\label{eq:Mori_projector}
\end{equation}
where the equilibrium inner product is
\begin{equation}
(A,B) \equiv \langle A^\dagger B\rangle_{\rm eq},
\label{eq:inner_product}
\end{equation}
and the average is taken in the equilibrium ensemble of the full microscopic system.
The complementary projector is
\begin{equation}
{\cal Q}=1-{\cal P}.
\label{eq:Q_projector}
\end{equation}

The exact Mori--Zwanzig identity then yields the generalized Langevin equation
for the vector \(\mathbf{A}(t)\):
\begin{equation}
\frac{d}{dt}\mathbf{A}(t)
=
i\boldsymbol{\Omega}\,\mathbf{A}(t)
-
\int_0^t ds\,
\mathbf{K}(s)\,\mathbf{A}(t-s)
+
\mathbf{F}(t),
\label{eq:MZ_vector}
\end{equation}
where the frequency matrix is
\begin{equation}
i\boldsymbol{\Omega}
=
(i{\cal L}\mathbf{A},\mathbf{A})\,(\mathbf{A},\mathbf{A})^{-1},
\label{eq:frequency_matrix}
\end{equation}
the fluctuating force is
\begin{equation}
\mathbf{F}(t)
=
e^{i{\cal Q}{\cal L}t}\,
{\cal Q}i{\cal L}\mathbf{A},
\label{eq:random_force_vector}
\end{equation}
and the memory kernel matrix is
\begin{equation}
\mathbf{K}(t)
=
(\mathbf{F}(0),\mathbf{F}(t))\,(\mathbf{A},\mathbf{A})^{-1}.
\label{eq:memory_matrix}
\end{equation}

Equation~\eqref{eq:MZ_vector} is exact. 

We now add a short derivation before the component equations and have also cited  the standard Mori--Zwanzig references already used in the paper.

The starting point is the standard Mori--Zwanzig equation for the vector of slow variables
\begin{equation}
A(t)
=
\left(
X(t),P(t),Q(t)
\right)^{T}.
\end{equation}

 What followed above was the standard Mori--Zwanzig procedure \cite{Zwanzig_book2001, Bagchi_book2023}. 

We next write component-wise for \(\mathbf{A}=(X,P,Q)^T\), 
to get

\begin{equation}
\begin{aligned}
\dot X(t) &= i\Omega_{XX}X(t)+i\Omega_{XP}P(t)+i\Omega_{XQ}Q(t) \\
&\quad - \int_0^t ds\,\Big[
K_{XX}(s)X(t-s)+K_{XP}(s)P(t-s) \\
&\qquad\qquad\qquad\qquad + K_{XQ}(s)Q(t-s)
\Big] + F_X(t)
\end{aligned}
\label{eq:MZ_X_general}
\end{equation}

Similarly,

\begin{equation}
\begin{split}
\dot P(t) &= i\Omega_{PX}X(t)
+i\Omega_{PP}P(t)
+i\Omega_{PQ}Q(t) \\
&\quad - \int_0^t ds\,\Big[
K_{PX}(s)X(t-s)+K_{PP}(s)P(t-s) \\
&\qquad\qquad\qquad\qquad + K_{PQ}(s)Q(t-s)
\Big] \\
&\quad + F_P(t)
\end{split}
\label{eq:MZ_P_general}
\end{equation}
And,

\begin{equation}
\begin{split}
\dot Q(t) &= i\Omega_{QX}X(t)
+i\Omega_{QP}P(t)
+i\Omega_{QQ}Q(t) \\
&\quad - \int_0^t ds\,\Big[
K_{QX}(s)X(t-s)+K_{QP}(s)P(t-s) \\
&\qquad\qquad\qquad\qquad + K_{QQ}(s)Q(t-s)
\Big] \\
&\quad + F_Q(t)
\end{split}
\label{eq:MZ_Q_general}
\end{equation}

For the present problem, the variable \(X\) is even under time reversal,
\(P\) is odd, and \(Q\) is naturally taken to be even. Under these symmetry conditions,
the reversible couplings simplify. In particular, one has

\begin{equation}
\dot X(t)=\frac{P(t)}{M},
\label{eq:Xdot_simple}
\end{equation}
so that Eq.~\eqref{eq:MZ_X_general} reduces to the well-known kinematic identity (when the effective
mass is independent of X).
%

The nontrivial projected dynamics is therefore carried by the coupled equations for
\(P(t)\) and \(Q(t)\). A convenient form is
\begin{equation}
\dot P(t)
=
-\frac{\partial W(X,Q)}{\partial X}
-\int_0^t ds\,
\Big[
K_{PP}(s)\,\frac{P(t-s)}{M}
+
K_{PQ}(s)\,Q(t-s)
\Big]
+F_P(t),
\label{eq:P_eq_coupled}
\end{equation}
\begin{equation}
\dot Q(t)
=
-\frac{\partial W(X,Q)}{\partial Q}
-\int_0^t ds\,
\Big[
K_{QP}(s)\,\frac{P(t-s)}{M}
+
K_{QQ}(s)\,Q(t-s)
\Big]
+F_Q(t).
\label{eq:Q_eq_coupled}
\end{equation}

Here \(W(X,Q)\) is the projected free-energy surface (or potential of mean force)
for the two slow variables, defined by
\begin{equation}
W(X,Q)
=
-k_BT\,
\ln \rho_{\rm eq}(X,Q)
+{\rm const},
\label{eq:W_def}
\end{equation}
where
\begin{equation}
\rho_{\rm eq}(X,Q)
\propto
\int d\Gamma\,
\delta\!\big(X-X(\Gamma)\big)\,
\delta\!\big(Q-Q(\Gamma)\big)\,
e^{-\beta H(\Gamma)}.
\label{eq:rho_XQ_def}
\end{equation}

Equation~\eqref{eq:W_def} is the two-dimensional generalization of the one-dimensional
entropic potential. If the neck geometry is described by an effective cross section
\(A(X,Q)\), then one may write
\begin{equation}
W(X,Q)
=
-k_BT\ln A(X,Q)+U_Q(Q)+W_{\rm other}(X,Q),
\label{eq:W_A_XQ}
\end{equation}
where \(U_Q(Q)\) is the intrinsic free-energy cost associated with the neck variable and
\(W_{\rm other}\) denotes any additional energetic or entropic contributions.

The fluctuating forces \(F_P(t)\) and \(F_Q(t)\) satisfy orthogonality relations
with the relevant variables,
\begin{equation}
(F_\mu(t),A_\nu)=0,
\qquad
\mu,\nu\in\{X,P,Q\},
\label{eq:orthogonality_random_force}
\end{equation}
and their correlations determine the memory kernels through the fluctuation--dissipation
structure,
\begin{equation}
K_{\mu\nu}(t)
=
\sum_{\lambda}
\big(F_\mu(0),F_\lambda(t)\big)
\big[(\mathbf{A},\mathbf{A})^{-1}\big]_{\lambda\nu}.
\label{eq:FDT_matrix}
\end{equation}

At this stage, Eqs. \eqref{eq:P_eq_coupled}, and
\eqref{eq:Q_eq_coupled} constitute the exact coupled Mori--Zwanzig description
for the progress coordinate and the neck variable \(Q\). Note that $P= M\dot X$.

For subsequent analysis it is useful to linearize the projected free-energy surface
near a representative bottleneck region. Retaining terms up to quadratic order gives
\begin{equation}
W(X,Q)
\simeq
W_0
+\frac{k_X}{2}X^2
+\frac{k_Q}{2}Q^2
-g\,XQ,
\label{eq:W_quadratic}
\end{equation}
where \(k_X\) is the local curvature along the progress coordinate, \(k_Q\) is the
stiffness associated with neck opening/closure, and \(g\) measures the coupling between
advance along \(X\) and the state of the neck.

The corresponding deterministic forces are then
\begin{equation}
\frac{\partial W}{\partial X}=k_X X-gQ,
\label{eq:dW_dX_linear}
\end{equation}
\begin{equation}
\frac{\partial W}{\partial Q}=k_Q Q-gX.
\label{eq:dW_dQ_linear}
\end{equation}

We substitute Eqs.~\eqref{eq:dW_dX_linear} and \eqref{eq:dW_dQ_linear}
into Eqs.~\eqref{eq:P_eq_coupled} and \eqref{eq:Q_eq_coupled}, we obtain the
linearized coupled generalized Langevin equations

\begin{equation}
\begin{split}
\dot P(t) &= -k_X X(t) + g Q(t) \\
&\quad - \int_0^t ds\,\Big[
K_{PP}(s)\frac{P(t-s)}{M} \\
&\qquad\qquad\qquad\qquad + K_{PQ}(s) Q(t-s)
\Big] \\
&\quad + F_P(t)
\end{split}
\end{equation}

And,

\begin{equation}
\begin{split}
\dot Q(t) &= -k_Q Q(t) + g X(t) \\
&\quad - \int_0^t ds\,\Big[
K_{QP}(s)\frac{P(t-s)}{M} \\
&\qquad\qquad\qquad\qquad + K_{QQ}(s) Q(t-s)
\Big] \\
&\quad + F_Q(t)
\end{split}
\end{equation}

The above two equations provide the minimal
two-variable Mori--Zwanzig framework for the entropic-neck problem. The slowing down
of motion along \(X\) is now generated self-consistently through its coupling to the
slow neck variable \(Q\). In the adiabatic limit where \(Q\) relaxes rapidly, one
recovers an effectively one-dimensional description for \(X\); when \(Q\) relaxes
slowly, elimination of \(Q\) produces an additional retarded friction kernel for
the motion along \(X\).

\subsection{Elimination of the neck variable and the induced friction kernel}

What follows is an established procedure,
although it is implemented here in the present context for the first time. The coupled Mori--Zwanzig equations derived above show that the neck variable \(Q\)
acts as an additional slow coordinate coupled to the progress variable \(X\).
To obtain a closed generalized Langevin equation for \(X\), we need to eliminate \(Q\).

if the neck variable relaxes rapidly compared with the progress coordinate, it may be treated as adiabatically adjusted to the instantaneous value of the progress coordinate. In that limit, the neck contributes mainly through the projected free-energy surface. When the neck relaxes slowly, however, eliminating it produces a retarded memory kernel.
This allows a useful simplification is to separate the reversible and dissipative effects of the neck.

To isolate this latter effect, it is convenient to work with the
velocity,
\begin{equation}
V(t) \equiv \dot X(t) = \frac{P(t)}{M},
\label{eq:V_def}
\end{equation}
and to retain explicitly the dynamical coupling between \(V\) and \(Q\).

We therefore consider the linearized reduced equations
\begin{equation}
\dot X(t)=V(t),
\label{eq:Xdot_reduced}
\end{equation}
\begin{equation}
M\dot V(t)
=
-k_X X(t)
-\int_0^t ds\,\Gamma_{\rm b}(t-s)\,V(s)
-\lambda Q(t)
+\eta_X(t),
\label{eq:Veq_reduced}
\end{equation}
\begin{equation}
\dot Q(t)
=
-\omega_Q Q(t)
-\int_0^t ds\,M_Q(t-s)\,Q(s)
+c\,V(t)
+\eta_Q(t).
\label{eq:Qeq_reduced}
\end{equation}

Here \(\Gamma_b(t)\) is the intrabasin memory kernel associated with fast local motions,
\(M_Q(t)\) is the memory kernel governing the relaxation of the neck variable itself,
and \(\omega_Q\) is the local harmonic restoring rate (equivalently, the bare inverse
timescale) for the relaxation of the neck coordinate \(Q\).
The parameters \(\lambda\) and \(c\) measure the strength of the coupling between the
neck and the motion along \(X\).

 The noises \(\eta_X(t)\) and \(\eta_Q(t)\) are the
projected fluctuating forces acting on \(V\) and \(Q\), respectively.

Equation~\eqref{eq:Qeq_reduced} is linear and may be solved exactly by Laplace
transformation. Defining the Laplace transform by
\begin{equation}
\tilde f(z)=\int_0^\infty dt\,e^{-zt}f(t),
\qquad
{\rm Re}\,z>0,
\label{eq:Laplace_def}
\end{equation}
we obtain from Eq.~\eqref{eq:Qeq_reduced}
\begin{equation}
z\tilde Q(z)-Q(0)
=
-\omega_Q \tilde Q(z)-\tilde M_Q(z)\tilde Q(z)
+c\,\tilde V(z)+\tilde \eta_Q(z),
\label{eq:Q_laplace_1}
\end{equation}
or
\begin{equation}
\tilde Q(z)
=
\tilde G_Q(z)\,
\Big[
Q(0)+c\,\tilde V(z)+\tilde \eta_Q(z)
\Big],
\label{eq:Q_laplace_solution}
\end{equation}
with the neck propagator
\begin{equation}
\tilde G_Q(z)
=
\frac{1}{z+\omega_Q+\tilde M_Q(z)}.
\label{eq:GQ_laplace}
\end{equation}

In the time domain, Eq.~\eqref{eq:Q_laplace_solution} becomes
\begin{equation}
Q(t)
=
G_Q(t)\,Q(0)
+
c\int_0^t ds\,G_Q(t-s)\,V(s)
+
\int_0^t ds\,G_Q(t-s)\,\eta_Q(s),
\label{eq:Q_time_solution}
\end{equation}
where \(G_Q(t)\) is the inverse Laplace transform of \(\tilde G_Q(z)\).

Equation~\eqref{eq:Q_time_solution} is the desired closed-form solution for the
neck variable: it shows that \(Q(t)\) contains three parts, namely
(i) propagation of the initial neck state,
(ii) the delayed response to the system velocity \(V(t)\), and
(iii) a fluctuating contribution arising from the hidden orthogonal dynamics.

Substituting Eq.~\eqref{eq:Q_time_solution} into Eq.~\eqref{eq:Veq_reduced},
we obtain
\begin{align}
M\dot V(t)
=&
-k_X X(t)
-\int_0^t ds\,\Gamma_{\rm b}(t-s)\,V(s)
\nonumber\\
&\;
-\lambda G_Q(t)\,Q(0)
-\lambda c \int_0^t ds\,G_Q(t-s)\,V(s)
-\lambda \int_0^t ds\,G_Q(t-s)\,\eta_Q(s)
+\eta_X(t).
\label{eq:V_after_Q_elim}
\end{align}

Collecting terms, the effective one-variable generalized Langevin equation for
\(X(t)\) becomes
\begin{equation}
M\ddot X(t)
=
-k_X X(t)
-\int_0^t ds\,\Gamma_{\rm eff}(t-s)\,\dot X(s)
+\Xi(t),
\label{eq:GLE_X_final}
\end{equation}
with the effective memory kernel
\begin{equation}
\Gamma_{\rm eff}(t)
=
\Gamma_{\rm b}(t)+\Gamma_{\rm neck}(t),
\label{eq:Gamma_eff_split}
\end{equation}
where the neck-induced contribution is
\begin{equation}
\Gamma_{\rm neck}(t)
=
\lambda c\,G_Q(t).
\label{eq:Gamma_neck_general}
\end{equation}

The effective fluctuating force is
\begin{equation}
\Xi(t)
=
\eta_X(t)
-\lambda \int_0^t ds\,G_Q(t-s)\,\eta_Q(s)
-\lambda G_Q(t)\,Q(0).
\label{eq:Xi_eff}
\end{equation}

The effective generalized Langevin equation obtained above retains the
fluctuation--dissipation structure inherited from the underlying
Mori--Zwanzig projection formalism. In particular, the effective random
force \(\Xi(t)\) is orthogonal to the retained slow variable, and its
autocorrelation determines the dissipative kernel. Thus, in equilibrium,
\begin{equation}
\langle \Xi(0)\,\Xi(t)\rangle = k_B T\,\Gamma_{\rm eff}(t),
\label{eq:FDT_Xi}
\end{equation}
up to normalization convention. Since
\(\Gamma_{\rm eff}(t)=\Gamma_{\rm b}(t)+\Gamma_{\rm neck}(t)\), the
noise correlation likewise contains both a fast intrabasin contribution
and a slow neck-induced contribution. This shows that the same neck
fluctuations that generate the long-time memory also generate the
corresponding fluctuating force.

A central result is that the retarded friction
kernel acting on the progress coordinate \(X\) is directly proportional to the
relaxation propagator of the neck variable \(Q\). 
Thus the memory kernel does not
have to be postulated; it emerges upon elimination of the fluctuating neck coordinate.

It is useful to consider a simple Markovian approximation for the neck dynamics.
If the internal neck kernel is taken to be local in time,
\begin{equation}
M_Q(t)=2\gamma_Q\,\delta(t),
\label{eq:MQ_markov}
\end{equation}

then Eq.~\eqref{eq:GQ_laplace} gives
\begin{equation}
\tilde G_Q(z)=\frac{1}{z+\omega_Q+\gamma_Q},
\label{eq:GQ_markov_laplace}
\end{equation}
so that
\begin{equation}
G_Q(t)=e^{-t/\tau_Q},
\qquad
\tau_Q^{-1}=\omega_Q+\gamma_Q.
\label{eq:GQ_markov_time}
\end{equation}
The neck-induced memory kernel then takes the simple exponential form
\begin{equation}
\Gamma_{\rm neck}(t)
=
\lambda c\,e^{-t/\tau_Q}.
\label{eq:Gamma_neck_exp}
\end{equation}

It is useful to emphasize the status of the result obtained so far. 
The Mori--Zwanzig elimination of the neck coordinate $Q$ gives a dynamical result:
the fluctuating entropic neck produces an additional retarded memory contribution
to the progress coordinate $X$. Thus the slow part of the friction is not postulated,
but arises from eliminating the hidden neck variable.

\textit{However, Eqs.~(64) and (65) do not by themselves determine the full temperature
dependence of the relaxation time.} To obtain the Adam--Gibbs form, one must further
relate the rarity of viable neck configurations to the configurational entropy. This
additional step is the cooperative-connectivity argument developed below. Thus the
Mori--Zwanzig construction provides the memory-kernel mechanism, while the
Adam--Gibbs dependence follows after the statistics of viable configurational necks
are connected to $s_{\rm c}(T)$.

The corresponding zero-frequency neck friction is
\begin{equation}
\zeta_{\rm neck}
=
\int_0^\infty dt\,\Gamma_{\rm neck}(t)
=
\lambda c\,\tau_Q.
\label{eq:zeta_neck}
\end{equation}

In general, the relaxation of the neck variable is non-Markovian, and the
memory kernel \(M_Q(t)\) cannot be reduced to a simple friction constant.
Only in the Markovian limit does \(\tilde M_Q(z)\) reduce to a constant
\(\gamma_Q\), leading to \(\tilde G_Q(z) = [z+\omega_Q+\gamma_Q]^{-1}\).

Equations~\eqref{eq:GLE_X_final}--\eqref{eq:zeta_neck} provide the desired
one-coordinate description of motion through the entropic neck. The fast intrabasin
motions contribute through \(\Gamma_{\rm b}(t)\), while the slow opening and closing
of the neck contributes the additional retarded kernel \(\Gamma_{\rm neck}(t)\).
When the neck relaxes rapidly, \(\tau_Q\) is small and the induced friction is weak
and short-lived. When the neck relaxes slowly, the induced kernel becomes long-lived,
and the motion along \(X\) is strongly retarded. In this sense, the slowdown of the
system emerges self-consistently from coupling to the hidden fluctuating neck variable.
%
\section{Neck-controlled memory, configurational entropy, and the Adam--Gibbs relation}

Unlike in Zwanzig-Fick-Jacobs description, our neck is not static. This makes the present problem
belong to a broad class of dynamical disorder model. We can describe these as "fluctuating bottle-neck"
models. The elimination of the neck variable \(Q\) within 
the Mori--Zwanzig
framework leads to an effective generalized Langevin equation for the
progress coordinate \(X\), given in Eq.~\eqref{eq:GLE_X_final}.
The corresponding memory kernel separates naturally into two contributions,
\begin{equation}
\Gamma_{\rm eff}(t)=\Gamma_{\rm b}(t)+\Gamma_{\rm neck}(t),
\qquad
\Gamma_{\rm neck}(t)=\lambda c\,G_Q(t),
\label{eq:Gamma_split_AG}
\end{equation}
where \(G_Q(t)\) is the relaxation propagator of the neck variable.
The term \(\Gamma_{\rm b}(t)\) describes fast intrabasin dynamics, while
\(\Gamma_{\rm neck}(t)\) encodes the slow dynamics associated with the
formation and relaxation of configurational bottlenecks.

The long-time dynamics is governed by the zero-frequency friction,
\begin{equation}
\zeta(T)=\int_0^\infty dt\,\Gamma_{\rm eff}(t)
=\zeta_{\rm b}(T)+\zeta_{\rm neck}(T),
\label{eq:GLE_AG_3}
\end{equation}
with
\begin{equation}
\zeta_{\rm neck}(T)=\lambda c\int_0^\infty dt\,G_Q(t).
\label{eq:GLE_AG_4}
\end{equation}
If the neck propagator is characterized by a dominant timescale \(\tau_Q(T)\),
then \(\zeta_{\rm neck}(T)\propto \tau_Q(T)\). In the regime where the
neck contribution dominates, the structural relaxation time follows
\begin{equation}
\tau_\alpha(T)\propto \tau_Q(T).
\label{eq:GLE_AG_5}
\end{equation}

The key physical step is to relate the timescale \(\tau_Q(T)\) to the rarity
of viable inter-basin connections in configuration space. The neck variable
\(Q\) should not be interpreted as a purely mechanical opening coordinate;
rather, it represents a coarse dynamical measure of configurational
connectivity between neighboring inherent-structure basins. A successful
relaxation event requires the system to access a subset of configurations
that both provide sufficient geometric opening and persist long enough
for passage. The probability of such viable neck configurations is therefore
controlled by their configurational measure.

Denoting by \(\Delta \Sigma_{\rm neck}(T)\) the entropy deficit associated
with these rare configurations, the mean waiting time for their occurrence
scales as
\begin{equation}
\tau_Q(T)=\tau_0 \exp\!\left[\Delta \Sigma_{\rm neck}(T)\right].
\label{eq:GLE_AG_6}
\end{equation}
This expression is the dynamical realization of Zwanzig's picture of diffusion
through entropic bottlenecks, in which the slowdown arises from the scarcity
of viable pathways in configuration space.

To connect this entropy deficit to thermodynamics, we identify the neck
configurations with the critical configurations that separate adjacent
metastable basins. In a free-energy description, these configurations are
associated with a barrier \(\Delta F_Q(T)\), so that
\begin{equation}
\Delta \Sigma_{\rm neck}(T)
=
\frac{\Delta F_Q(T)}{k_B T}.
\label{eq:GLE_AG_7}
\end{equation}
This identification establishes the equivalence between the entropic-neck
description and the free-energy barrier picture: the rarity of bottleneck
configurations is directly related to the activation free energy.

We now add the configurational-entropy input that is not contained in the
Mori--Zwanzig elimination itself. The memory-kernel analysis shows that the
slow relaxation is controlled by the lifetime and recurrence of viable neck
configurations. To obtain the temperature dependence of this slow process, we
must estimate how rare such viable necks are in the landscape.

Let $s_{\rm c}(T)$ denote the configurational entropy per particle in units of
$k_{\rm B}$. A region containing $z$ particles has, up to subexponential factors,
a number of distinct amorphous configurations of order
\begin{equation}
{\cal N}(z,T)
\sim
\exp
\left[
z s_{\rm c}(T)
\right].
\label{eq:Nconfig_z}
\end{equation}

Let $(S_c(T)$ denote the total configurational entropy and
$(s_c(T)$ the configurational entropy per particle measured in units of $k_B$. 

For a viable interbasin passage to exist, this region must contain at least one
configuration that opens a sufficiently persistent neck. We represent the
logarithmic measure required for such connectivity by a dimensionless threshold
$\Delta\Sigma_c$. The minimal cooperative size is then determined by
\begin{equation}
z^{*}(T)s_{\rm c}(T)
\simeq
\Delta\Sigma_c .
\label{eq:zstar_connectivity}
\end{equation}
Thus
\begin{equation}
z^{*}(T)
=
\frac{\Delta\Sigma_c}
{s_{\rm c}(T)} .
\label{eq:zstar_sc}
\end{equation}
This is the cooperative-connectivity step. It expresses the simple fact that, as
the configurational entropy decreases on cooling, a larger region must rearrange
coherently in order to provide a viable pathway between basins.

This is the cooperative-connectivity step. It expresses the simple fact that,
as the configurational entropy decreases on cooling, a larger region must
rearrange coherently in order to provide a viable pathway between basins.
If the activation cost for creating or traversing such a viable neck is
approximately proportional to the number of participating degrees of freedom,
then
\begin{equation}
\Delta F^{\ddagger}(T)
=
\mu z^{*}(T)
=
\frac{\mu\Delta\Sigma_c}{s_c(T)} ,
\end{equation}
where \(\mu\) is an effective free-energy cost per participating unit.

If the activation cost for creating or traversing such a viable neck is
approximately proportional to the number of participating degrees of freedom,
then
\begin{equation}
\Delta F^{\ddagger}(T)
=
\mu z^{*}(T)
=
\frac{\mu\Delta\Sigma_c}
{s_{\rm c}(T)} ,
\label{eq:barrier_zstar}
\end{equation}

Substitution
into the activated form of the relaxation time gives
\begin{equation}
\tau_{\alpha}(T)
=
\tau_0
\exp
\left[
\frac{\Delta F^{\ddagger}(T)}
{k_{\rm B}T}
\right]
=
\tau_0
\exp
\left[
\frac{C}
{T s_{\rm c}(T)}
\right],
\label{eq:AG_from_connectivity}
\end{equation}
where $C=\mu\Delta\Sigma_c/k_{\rm B}$. This is the Adam--Gibbs form. Thus the Adam--Gibbs dependence is not obtained from the projection formalism alone; it
appears when the neck-controlled memory kernel is combined with the
cooperative-connectivity estimate for the rarity of viable configurational necks.

We now compare the present entropic-neck model with the Random First-Order Transition (RFOT) framework, in which the free-energy cost of forming a viable cooperative rearrangement is governed by the competition between interfacial mismatch and configurational entropy, leading to
\begin{equation}
\Delta F_Q(T)\propto \frac{1}{s_c(T)},
\label{eq:GLE_AG_8}
\end{equation}
where \(s_c(T)\) is the configurational entropy per particle. Substituting
Eqs.~\eqref{eq:GLE_AG_7} and \eqref{eq:GLE_AG_8} into
Eq.~\eqref{eq:GLE_AG_6}, and using Eq.~\eqref{eq:GLE_AG_5}, we obtain

\begin{equation}
\tau_\alpha(T)
\sim
\tau_0\exp\!\left[\frac{C}{T s_c(T)}\right],
\label{eq:GLE_AG_9}
\end{equation}
which is the Adam--Gibbs relation.

Thus, within the present Mori--Zwanzig framework, the Adam--Gibbs law emerges
as a consequence of neck-controlled memory: the slow relaxation kernel is
governed by the dynamics of rare configurational bottlenecks, whose entropy
deficit is equivalent to a free-energy barrier that scales inversely with
the configurational entropy.

In this sense, the Zwanzig entropic-neck picture
and the RFOT nucleation picture provide complementary descriptions of the
same underlying bottleneck controlling structural relaxation.

The above discussion clarifies the status of the Adam--Gibbs result in the present theory. 
The Mori--Zwanzig formulation establishes the dynamical mechanism: elimination of the 
fluctuating neck variable produces a slow memory kernel for the progress coordinate. 
The Adam--Gibbs dependence follows only after this dynamical result is combined with 
the cooperative-connectivity condition, which relates the rarity of viable necks to the 
configurational entropy $s_{\rm c}(T)$. Thus the present approach does not replace the 
Adam--Gibbs argument, but gives it a geometrical and memory-function interpretation 
in terms of fluctuating entropic bottlenecks.

\section{Connection counting, viable pathways, and the origin of Adam--Gibbs scaling}

A useful way to understand the emergence of activated dynamics is to
separate the total number of inter-basin connections from the fraction
of those that are dynamically viable.

At high temperature the effective neck is broad, and the dominant relaxation channel is local intrabasin motion. This motion is controlled mainly by short-range structure and is therefore naturally correlated with the excess entropy, giving Rosenfeld-type scaling. Upon cooling, the local channel remains present, but the slow interbasin neck channel grows and eventually dominates the long-time friction.

Let us elaborate. The Mori--Zwanzig analysis shows that, in the neck-dominated regime, the
structural relaxation time is controlled by the timescale \(\tau_Q(T)\)
associated with the fluctuating neck variable \(Q\). To make contact with
the configurational interpretation of this timescale, it is useful to view
relaxation as a two-step process. First, the system must possess a set of
possible inter-basin connections determined by the configurational entropy
of the basin. Second, among these possible connections, only a small subset
will be dynamically viable, i.e., sufficiently open and sufficiently persistent
to permit passage. The overall relaxation rate is therefore naturally written
as the product of the number of available connections and the probability
that a given connection is viable.

We therefore write the rate of relaxation as
\begin{equation}
k(T) \sim k_0\, N_{\rm connections}(T)\, P_{\rm viable}(T),
\label{eq:k_factorized}
\end{equation}
where \(N_{\rm connections}\) is the total number of configurational
pathways leading out of a given basin, and \(P_{\rm viable}\) is the
probability that a given pathway is sufficiently open and persistent to
allow relaxation.

The total number of connections is controlled by the configurational
entropy. If \(s_c(T)\) is the configurational entropy per particle, then
\begin{equation}
N_{\rm connections}(T) \sim \exp\!\left[\frac{s_c(T)}{k_B}\right],
\label{eq:N_connections}
\end{equation}
which reflects the exponential growth of available configurations with
entropy. This contribution alone leads to an increase of the rate with
entropy and corresponds to the high-temperature, Rosenfeld-type regime.

At low temperatures, however, not all connections are dynamically
accessible. A viable pathway requires a cooperative rearrangement of a
region of size \(z^\ast(T)\), such that the neck is both wide enough and
remains open long enough for passage. The probability of such a
cooperative event is exponentially small,
\begin{equation}
P_{\rm viable}(T)
\sim
\exp\!\left[-\frac{\Delta \mu}{k_B T}\, z^\ast(T)\right],
\label{eq:P_viable}
\end{equation}
where \(\Delta \mu\) is an effective free-energy cost per particle.

The cooperative size \(z^\ast(T)\) is determined by the condition that
at least one viable pathway exists within the available configurational
space. Since the number of distinct configurations of a region of size
\(z\) scales as \(\exp[z\,s_c(T)/k_B]\), this condition yields
\begin{equation}
z^\ast(T)\, \frac{s_c(T)}{k_B} \sim \mathcal{O}(1),
\qquad
\Rightarrow
\qquad
z^\ast(T) \propto \frac{k_B}{s_c(T)}.
\label{eq:zstar_sc}
\end{equation}

Substituting Eq.~\eqref{eq:zstar_sc} into Eq.~\eqref{eq:P_viable}, we obtain
\begin{equation}
P_{\rm viable}(T)
\sim
\exp\!\left[-\frac{C}{T s_c(T)}\right],
\label{eq:P_viable_AG}
\end{equation}
with \(C\) a constant. Combining Eqs.~\eqref{eq:k_factorized},
\eqref{eq:N_connections}, and \eqref{eq:P_viable_AG}, we find
\begin{equation}
k(T)
\sim
k_0 \exp\!\left[\frac{s_c(T)}{k_B}\right]
\exp\!\left[-\frac{C}{T s_c(T)}\right].
\label{eq:k_full}
\end{equation}
We note that although \eqref{eq:k_full} has the appearance of an extrapolation equation (although a non-trivial one), this has been derived here employing the Mori-Zwanzig formalism, and is fairly robust.

The present theoretical formulation explains the crossover
from Rosenfeld scaling to Adam--Gibbs behavior as a change in the
entropy that controls the dominant relaxation mechanism.
In particular, the excess entropy can be decomposed as \(s_{\rm ex}=s_c+s_{\rm vib}\), but while the vibrational entropy \(s_{\rm vib}\) contributes to local structural fluctuations, it does not facilitate inter-basin transitions; hence it drops out of the activated dynamics, leaving the configurational entropy \(s_c\) as the relevant control parameter at low temperatures.

At high temperatures, the first factor dominates and the rate increases
with entropy. Upon supercooling, the second factor becomes dominant,
leading to
\begin{equation}
\tau(T) \sim k(T)^{-1}
\sim
\tau_0 \exp\!\left[\frac{C}{T s_c(T)}\right],
\label{eq:AG_final}
\end{equation}
which is the Adam--Gibbs relation.

This decomposition makes clear that the configurational entropy plays a
dual role: it increases the number of available pathways, but at the same
time controls the size of the cooperative region required for a viable
transition. The latter effect dominates at low temperatures and gives
rise to activated, entropy-controlled dynamics.

A noteworthy aspect of the above result is the appearance of the
configurational entropy \(s_c(T)\), whereas the high-temperature
Rosenfeld relation is expressed in terms of the excess entropy
\(s_{\rm ex}\). This difference reflects the change in the nature
of the dynamical bottleneck. At high temperatures, transport is
dominated by local collisional motion, and the relevant entropy is
the excess entropy, which captures short-range structural correlations.
Upon supercooling, the system becomes confined to metastable basins,
and relaxation is governed by rare transitions between basins.\cite{StillingerWeber1,StillingerWeber2}

The two entropic contributions are controlled by different entropy measures because they describe different kinds of motion. The Rosenfeld channel is local and intrabasin; it involves collisional motion, cage exploration, and short-range structural rearrangements, and is therefore naturally correlated with the excess entropy $S_{\rm ex}$. The neck channel, by contrast, describes interbasin escape through rare viable bottlenecks. Its probability is controlled by the number of distinct amorphous basins available to the rearranging region, and hence by the configurational entropy $S_{\rm c}$. Thus the crossover from Rosenfeld to Adam--Gibbs behavior is also a crossover from an $S_{\rm ex}$-controlled local channel to an $S_{\rm c}$-controlled activated channel.

However, there is a caveat in the above logic. Simulations  have emphasized both the usefulness and the limitations of inherent-structure-based estimates. Such estimates often capture the qualitative decrease of configurational entropy upon supercooling, but they need not coincide quantitatively with more rigorous definitions used in the RFOT context.

\textit{The corresponding bottleneck is therefore controlled by the number
of accessible basins, i.e., the configurational entropy \(s_c(T)\).}
In this sense, the present formulation naturally explains the crossover
from Rosenfeld scaling to Adam--Gibbs behavior as a change in the
entropy that controls the dominant relaxation mechanism.
This has been partly addressed above and seems fairly straightforward now in view of all the analyses done above.
In this section we argue how the Adam--Gibbs (AG) relation could  emerge from 
the narrowing of an entropic neck in configuration space.
The central idea is that the entropy deficit associated with the neck,
\(\Delta S_{\rm neck}(T)\),
generates a temperature-dependent entropic barrier
\[
\Delta F_{\rm neck}(T) = T\,\Delta S_{\rm neck}(T),
\]
whose growth upon cooling yields super-Arrhenius behavior.
The speed with which this barrier increases with decreasing \(T\)---that is, the 
\emph{sharpness} of neck collapse---governs the fragility of the liquid.

\subsection{Entropic barrier from the neck entropy deficit}

As established in section 2, the entropy deficit associated with an 
entropic bottleneck may be written as
\[
\Delta S_{\rm neck}(T)
= S_{\rm basin}(T) - S_{\rm neck}(T),
\]
where \(S_{\rm basin}\) is the entropy available within the basin and 
\(S_{\rm neck}\) is the entropy of the constricted region through which the 
system must pass to escape.
A simple and physically transparent representation is
\begin{equation}
\Delta S_{\rm neck}(T)
= \lambda\, n_{\rm act}(T),
\label{eq:DSneck_basic}
\end{equation}
where \(\lambda\) is the entropic cost per active coordinate, and
\[
n_{\rm act}(T) = \phi(T)\,z^\star(T)
\]
is the number of degrees of freedom that must rearrange cooperatively.
Here \(\phi(T)\in(0,1]\) is the participation fraction and \(z^\star(T)\) is the 
minimal cooperative size required for maintaining configurational connectivity.

The corresponding free-energy barrier is therefore
\begin{equation}
\Delta F_{\rm neck}(T)
= T\,\lambda\,\phi(T)\,z^\star(T).
\label{eq:Fneck_basic}
\end{equation}

\subsection{Minimal cooperative size and its scaling with configurational entropy}

Following the connectivity argument introduced earlier, the minimal cooperative 
size obeys
\begin{equation}
z^\star(T) = \frac{\Delta\Sigma}{k_B\,s_c(T)},
\label{eq:zstar}
\end{equation}
where \(\Delta\Sigma\) represents a characteristic entropy deficit needed to open 
a viable pathway, and \(s_c(T)\) is the configurational entropy per particle.
Equation~\eqref{eq:zstar} expresses a simple but profound idea:
as the configurational entropy decreases, larger regions must reorganize 
coherently to maintain dynamical connectivity.
Combining Eqs.~\eqref{eq:DSneck_basic} and \eqref{eq:zstar}, we obtain
\begin{equation}
\Delta S_{\rm neck}(T)
= \lambda\,\phi(T)\,\frac{\Delta\Sigma}{k_B\,s_c(T)}.
\label{eq:DSneck_sc}
\end{equation}

Since \(s_c(T)\) decreases rapidly upon cooling in fragile liquids, 
\(\Delta S_{\rm neck}(T)\) grows sharply, generating a large entropic barrier 
that suppresses basin-to-basin transitions.

\subsection{Adam--Gibbs form from entropic-neck picture}
We substitute Eq.~\eqref{eq:DSneck_sc} into Eq.~\eqref{eq:Fneck_basic}, we obtain
\[
\Delta F_{\rm neck}(T)
= T\,\lambda\,\phi(T)\,
\frac{\Delta\Sigma}{k_B\,s_c(T)}.
\]
Assuming that the participation fraction \(\phi(T)\) varies slowly over
the temperature range of interest, the structural relaxation time can be
written in Eyring form as
\begin{equation}
\tau(T) = \tau_0
\exp\!\left[
\frac{\lambda\,\phi(T)\,\Delta\Sigma}{k_B\,T\,s_c(T)}
\right].
\label{eq:AG_neck_prefinal}
\end{equation}
Defining \(C=\lambda\,\phi(T)\,\Delta\Sigma/k_B\), we obtain
\begin{equation}
\tau(T)
= \tau_0
\exp\!\left[\,\frac{C}{T\,s_c(T)}\,\right],
\label{eq:AG_final}
\end{equation}
which is the Adam--Gibbs relation.

Thus, the AG form follows directly from the geometric picture of diffusion
through a narrowing entropic neck. The prefactor \(C\) reflects the
material-dependent entropy deficit \(\Delta\Sigma\) and the participation
of the active degrees of freedom.
\subsection{Rosenfeld scaling in the entropy neck picture}

At high temperatures the configurational entropy is large and 
\(s_c(T)\) varies slowly.
In this regime, \(z^\star(T)\) and therefore \(\Delta S_{\rm neck}(T)\) are 
small:
\[
\Delta S_{\rm neck}(T) \ll 1
\quad\Longrightarrow\quad
\Delta F_{\rm neck}(T) \sim 0.
\]
Thus relaxation is not activation-limited but rather governed by local, 
collisional motion within a basin.
Transport coefficients are then controlled primarily by the excess entropy of 
the liquid, yielding Rosenfeld-type scaling relations such as
\[
D^* \sim \exp(\alpha S_{\rm ex}).
\]
The entropic neck is effectively ``open'' at high temperature; only when 
cooling drives \(s_c(T)\) downward does the neck constrict enough to generate 
a barrier.
\subsection {Crossover behavior}

Here we provide further clarification of the crossover behavior. The fast
Rosenfeld-like contribution may be written schematically as
\begin{equation}
k_R(T)=k_0\exp[\alpha s_{\rm ex}(T)] ,
\end{equation}
where \(s_{\rm ex}(T)\) is the excess entropy in units of \(k_B\). This contribution
describes the fast, local part of the dynamics and is distinct from the activated
neck-controlled channel that dominates at lower temperature.

Equivalently, in a friction representation, the total friction may be written as
\begin{equation}
\zeta(T)=\zeta_R(T)+\zeta_{\rm neck}(T).
\end{equation}

Here $\zeta_{\rm R}(T)$ is the local intrabasin contribution, while $\zeta_{\rm neck}(T)$ is the slow interbasin contribution arising from the fluctuating entropic neck. The neck contribution has the dominant activated dependence
\begin{equation}
\zeta_{\rm neck}(T)
\simeq
\zeta_0
\exp
\left[
\frac{C}
{T s_{\rm c}(T)}
\right].
\label{eq}
\end{equation}
The crossover temperature is then determined approximately by
\begin{equation}
\zeta_{\rm R}(T_{\rm cross})
\simeq
\zeta_{\rm neck}(T_{\rm cross}).
\label{eq}
\end{equation}

At high temperature, the dominant motion is local and intrabasin. It involves collisional dynamics, cage exploration, and short-range structural rearrangements. These processes are naturally correlated with the excess entropy $S_{\rm ex}$, or with its pair contribution in simple liquids. At lower temperature, however, the rate-limiting step is no longer local exploration within a basin, but escape from one metastable basin to another through a rare viable neck. The number of distinct amorphous basins available to the rearranging region is controlled by the configurational entropy $S_{\rm c}$. Thus the local Rosenfeld channel is governed by $S_{\rm ex}$, whereas the activated neck channel is governed by $S_{\rm c}$.

\subsection{Fragility as the rate of neck collapse}

The sharpness of the increase in relaxation times upon cooling---the 
\emph{fragility} of the liquid---is governed by the rate at which 
\(\Delta S_{\rm neck}(T)\) grows.
Differentiating Eq.~\eqref{eq:DSneck_sc}, we obtain
\[
\frac{d\Delta S_{\rm neck}}{dT}
= -\lambda\,\phi(T)\,\frac{\Delta\Sigma}{k_B}\,
\frac{1}{s_c(T)^2}\,\frac{ds_c}{dT},
\]
indicating that liquids in which \(s_c(T)\) decreases rapidly (large 
\(|ds_c/dT|\)) exhibit rapid growth of the entropic barrier.
This identifies the kinetic fragility parameter \(m\) with the slope of the neck 
entropy at the glass transition:
\begin{equation}
m = 
-\frac{T_g}{\ln 10}
\left.\frac{d\ln \tau}{dT}\right|_{T_g}
\propto
-\frac{T_g}{k_B}
\left.\frac{d\,\Delta S_{\rm neck}}{dT}\right|_{T_g}.
\label{eq:fragility_neck}
\end{equation}
Equation~\eqref{eq:fragility_neck} shows that fragile liquids correspond to 
\emph{rapid neck collapse}, while strong liquids have gradually narrowing necks.
This provides a microscopic interpretation of fragility within the entropic-neck 
framework.

\subsection {Bridging of MCT and RFOT and the crossover}

 In a detailed work, Bhattacharyya, Bagchi and Wolynes (BBW) developed a description that bridged the gap between 
 the mode coupling theory (MCT) with the random first order transition (RFOT) theory of glassy relaxation. 
 \cite{BBW_JCP_2004,BBW_PRE_2005,BBW_PNAS_2008}
 In their
 approach, the dynamic structure factor is assumed to be given by
 \begin{equation}
 S(k,\omega) = \frac {S(k)} {i\omega + M(k,\omega)},
 \end{equation}
 where,
 \begin {equation}
 M(k,\omega) = \Gamma_{MCT} (k,\omega) + \Gamma_{RFOT} (k,\omega)
 \end{equation}.

 That is, the memory function is the sum of two contributions. At this stage, 
 the two appears to be unrelated. However, the MCT memory function is to be calculated self-consistently
 using the full dynamic structure $S(k,\omega)$, and thus gets coupled to the 
 RFOT-hopping term \cite{BBW_JCP_2004,BBW_PRE_2005,BBW_PNAS_2008}.

The entropic-neck mechanism provides a unified description of 
Rosenfeld-type behavior at high temperatures and Adam--Gibbs behavior at low 
temperatures.
In the high-temperature regime the neck is wide, the entropy deficit is small, 
and dynamics correlate with excess entropy.
Upon cooling, the configurational entropy decreases, the neck constricts, and a 
growing entropic barrier emerges.
The AG relation follows directly from the dependence of the cooperative size on $(1/s_{c}(T)$).
The fragility of a liquid is determined by the rate at which the neck entropy 
grows: fragile liquids exhibit rapid neck collapse and sharp dynamic crossover, 
while strong liquids exhibit gradual changes and broad crossover.

The high-temperature regime is dominated by local intrabasin or collisional dynamics, and is therefore better described by excess entropy, or by its pair contribution in simple liquids. We reserve $s_{\rm c}(T)$ for the supercooled regime where metastable basins and interbasin transitions become meaningful.
%
\section{Parallel Memory-Function Decomposition of Dynamical Friction}
\label{sec:parallel_memory}

The entropic-neck mechanism developed in earlier sections 
implies that the structural relaxation of a supercooled liquid proceeds through  
\emph{two statistically distinct dynamical channels}:

\begin{enumerate}
\item[(i)] a high-entropy, intrabasin channel associated with 
local collisional motion and cage exploration, dominant at high temperatures;  

\item[(ii)] a low-entropy, interbasin channel associated with 
traversing the entropic neck, dominant in the deeply supercooled regime.
\end{enumerate}

These two channels act in \emph{parallel} on the generalized friction kernel
of the particle or collective mode of interest.
This section formulates this insight using the Mori--Zwanzig memory-function
framework and derives a transparent expression for the dynamical crossover.

\subsection{Additive decomposition of the memory kernel}

Let the generalized Langevin equation for a slow variable \(A(t)\) be
\begin{equation}
\dot{A}(t) = -\int_0^t M(t-t')\,A(t')\,dt' + \xi(t),
\label{eq:GLE_basic}
\end{equation}
with memory kernel \(M(t)\) and noise \(\xi(t)\) satisfying the fluctuation–
dissipation relation.
The essential observation is that the noise may be decomposed into statistically
independent parts,
\[
\xi(t) = \xi_R(t) + \xi_{\rm neck}(t),
\]
corresponding to local intrabasin fluctuations (Rosenfeld channel) and 
interbasin fluctuations mediated by neck passage (AG channel).
Statistical independence implies vanishing cross-correlation:
\[
\langle \xi_R(t)\,\xi_{\rm neck}(0)\rangle = 0.
\]
Therefore, by the Mori--Zwanzig prescription,
\begin{equation}
M(t) = M_R(t) + M_{\rm neck}(t).
\label{eq:kernel_additive}
\end{equation}

This additivity expresses the fact that the two channels contribute 
independently to the friction experienced by the slow degrees of freedom.
Importantly, the two contributions dominate in different temperature regimes:
\[
M_R(t) \gg M_{\rm neck}(t) \quad (T \gg T_{\rm cross}),
\]
\[
M_{\rm neck}(t) \gg M_R(t) \quad (T \ll T_{\rm cross}).
\]

\subsection{Long-time friction and the diffusion coefficient}

The effective (zero-frequency) friction is 
\[
\zeta(T) = \int_0^\infty M(t)\,dt
= \zeta_R(T) + \zeta_{\rm neck}(T),
\]
where
\[
\zeta_R(T) = \int_0^\infty M_R(t)\,dt,
\qquad
\zeta_{\rm neck}(T) = \int_0^\infty M_{\rm neck}(t)\,dt.
\]
The diffusion coefficient follows directly from the generalized Einstein relation,
\begin{equation}
D(T) = \frac{k_B T}{\zeta_R(T) + \zeta_{\rm neck}(T)}.
\label{eq:D_parallel}
\end{equation}

Equation~\eqref{eq:D_parallel} is the mathematical expression of the 
\emph{parallel-channel} picture:
the dominant friction channel at a given temperature controls the observed 
transport dynamics.


The Rosenfeld channel represents local, high-entropy dynamics.
Its memory kernel is short-ranged in time and weakly dependent on temperature:
\[
M_R(t) \sim \omega_0^2\,e^{-t/\tau_0},
\]
with 
\(\tau_0\sim 10^{-2}$--$10^{-1}\,{\rm ps}\)
at typical liquid densities.
The corresponding friction behaves as
\begin{equation}
\zeta_R(T)
\propto \exp[-\alpha\,S_{\rm ex}(T)],
\label{eq:zeta_R_RS}
\end{equation}
consistent with Rosenfeld scaling relations.
Thus,
\[
D_R(T) \propto \exp[\alpha\,S_{\rm ex}(T)],
\]
where \(D_R(T) = k_B T/\zeta_R(T)\) is the diffusion coefficient associated with the 
Rosenfeld channel.


In contrast, the neck channel captures transitions between basins and therefore
exhibits activated behavior.
From Sec.4, the associated relaxation time obeys
\[
\tau_{\rm neck}(T)
= \tau_0
\exp\!\left[\frac{C}{T\,s_c(T)}\right].
\]
Thus the long-time friction associated with the neck channel grows rapidly:
\begin{equation}
\zeta_{\rm neck}(T)
\sim \zeta_0\,\exp\!\left[\frac{C}{T\,s_c(T)}\right].
\label{eq:zeta_neck_AG}
\end{equation}
The prefactor \(\zeta_0\) depends on the short-time dynamics but varies slowly
with temperature compared to the exponential factor.

\subsection{Fragility dependence of crossover temperature }

The RS-to-AG dynamic crossover occurs when the two friction contributions become
comparable:
\[
\zeta_R(T_{\rm cross}) \approx \zeta_{\rm neck}(T_{\rm cross}).
\]
Substituting Eqs.~\eqref{eq:zeta_R_RS} and \eqref{eq:zeta_neck_AG}, one finds a 
transcendental equation for \(T_{\rm cross}\).
A useful approximate criterion is obtained by equating the arguments of the two
exponentials:
\begin{equation}
\alpha\,S_{\rm ex}(T_{\rm cross})
\approx 
\frac{C}{T_{\rm cross}\,s_c(T_{\rm cross})}.
\label{eq:cross_criterion}
\end{equation}

Strong liquids show gentle temperature dependence of both 
\(S_{\rm ex}(T)\) and \(s_c(T)\), leading to broad crossover.
Fragile liquids exhibit rapid collapse of configurational entropy 
and therefore satisfy the criterion more sharply, producing a narrow and 
distinct dynamic crossover.

Equation~\eqref{eq:cross_criterion} provides a quantitative means of comparing 
different materials and establishes the crossover as a 
\emph{fragility-dependent phenomenon}, governed by entropy competition between 
the two friction channels.


The parallel memory-function decomposition describes equilibrium dynamics.
Structural nonequilibrium---such as cooling or heating through \(T_g\)---requires
a third memory kernel \(K(t)\), 
governing the evolution of the fictive temperature \(T_f(t)\).
In a latter section  we  introduce this enthalpy kernel and show how 
the rate of fictive-temperature lag is governed by the same entropic-neck
mechanism that determines fragility and the equilibrium crossover.
%

\section { Connection with other well-known theories}
\label{sec:RFOT}

The entropic-neck framework developed in the preceding sections provides a 
geometric mechanism for the emergence of cooperative relaxation and the 
Adam--Gibbs relation.
On the other hand, the mathematical structure appears in the  
Random First-Order Transition (RFOT) theory, which is more quantitative and  where activated dynamics arise 
from a competition between configurational entropy and an interfacial mismatch 
penalty between amorphous states.
This section discusses that the entropic neck and the RFOT droplet correspond to 
two complementary descriptions of the same physical bottleneck.

\subsection{What probability is computed: neck creation versus neck traversal}

Although the RFOT droplet and the entropic neck lead to formally identical
Adam--Gibbs scaling of the relaxation time, they address \emph{distinct physical
questions} and compute \emph{different probabilities}.
Clarifying this distinction is essential for interpreting the correspondence
established above.

Within the Stillinger--Weber landscape picture \cite{StillingerWeber1,StillingerWeber2}, the instantaneous state of the
system corresponds to a point in a high-dimensional configuration space.
Metastable amorphous states occupy extended basins of measure
$\Omega_{\rm basin}$, while transitions between basins are possible only through
narrow regions---or necks---of much smaller measure $\Omega_{\rm neck}$.
Structural relaxation therefore requires configurational connectivity through
such necks.

In the \emph{entropic-neck} (Zwanzig--Fick--Jacobs) picture developed in this
work, the neck is treated as a \emph{rare but pre-existing} region of
configuration space.
The problem is kinetic rather than thermodynamic:
the system diffuses within a basin and must \emph{find} a viable neck and
\emph{pass through it} before it closes.
The relevant probability is therefore
\begin{equation}
P_{\rm pass}
\;\sim\;
\frac{\Omega_{\rm neck}}{\Omega_{\rm basin}},
\end{equation}
and the associated free-energy barrier is purely entropic,
\[
\Delta F_{\rm neck}
= -k_B T \ln\!\left(\frac{\Omega_{\rm neck}}{\Omega_{\rm basin}}\right).
\]
Here the basin carries the entropy, while the neck is entropically depleted.
Additional constraints---such as finite particle size, orientational alignment,
or collective compatibility---further reduce $\Omega_{\rm neck}$ and enhance the
barrier.
The Adam--Gibbs form emerges when $\Omega_{\rm neck}$ decreases exponentially
with the number of participating degrees of freedom.

RFOT, by contrast, addresses a different, and in our view, possibly a more plausible scenario for 
relaxation in a heterogeneous glassy liquid.
The system is assumed to be trapped in a metastable amorphous state, and
structural relaxation requires the \emph{creation} of a rearranging region (or
entropy droplet) of sufficient size.
The free-energy cost $\Delta F_{\rm RFOT}(N)$ represents the probability of
\emph{forming} a neck-like connection between two amorphous states,
\begin{equation}
P_{\rm form}
\;\sim\;
\exp[-\beta\,\Delta F_{\rm RFOT}(N^\star)].
\end{equation}
In this picture, the neck does not pre-exist; it is stabilized by the entropy
gain of the new configuration against the energetic penalty of surface
mismatch.
Thus RFOT computes the probability of \emph{creating} a neck, whereas the
entropic-neck picture computes the probability of \emph{finding and traversing}
one.

The two viewpoints may be reconciled dynamically by recognizing that, in a
fluctuating many-body system, configurational necks continuously appear and
disappear.
Structural relaxation then requires two conditions to be satisfied:
(i) a neck of sufficient size must form, and
(ii) the system must locate and pass through that neck during its lifetime.
Schematically,
\[
\tau^{-1}
\;\sim\;
P_{\rm form}\times P_{\rm pass}.
\]
The RFOT framework emphasizes condition (i), while the Zwanzig entropic-neck
framework emphasizes condition (ii).
They represent complementary projections of the same underlying process, with
different entropy bookkeeping:
in RFOT the entropy stabilizes the neck, whereas in the entropic-neck picture
the entropy resides primarily in the basin and is lost at the constriction.

This distinction explains why the two approaches yield identical scaling with
$s_c(T)$ and fragility while retaining distinct physical interpretations.

The two perspectives therefore represent different projections of the same 
underlying transition between amorphous states: one emphasizes an interfacial 
energy in real space, the other an entropic constriction in configuration space.

\section{Conclusion}

We have attempted to develop a theoretical framework to understand the anomalous crossover dynamics
in supercooled liquids based on the concept of an \emph{entropic neck} in
configuration space. In this picture, relaxation is controlled by the
formation and persistence of rare bottleneck configurations that connect
metastable basins. The central idea is that the slowdown of dynamics arises
from an entropy deficit associated with these viable pathways.

At the same time, it is important to recognize a limitation of the present
formulation. The entropic-neck description, in its simplest form, treats
relaxation as diffusion through an already existing bottleneck and thus
emphasizes the \emph{search for} viable pathways. In deeply supercooled or
glassy systems, however, such pathways are not generically available and
must themselves be created through cooperative rearrangements of many
degrees of freedom. In this regime, the formation of the neck becomes an
activated process, closely related to ideas underlying nucleation-based
approaches such as the RFOT framework. The present treatment captures the
statistics and dynamical consequences of these bottlenecks once they are
accessible, but does not explicitly describe the microscopic mechanism by
which they are generated.

This distinction clarifies the scope of the theory. The entropic-neck picture
provides a transparent and quantitatively tractable description of how the
scarcity and temporal persistence of viable pathways control relaxation
dynamics, and naturally yields the observed entropy-controlled relations.
However, a complete description of glassy relaxation must also account for
the cooperative processes responsible for the emergence of such pathways in
the first place. Establishing this connection more explicitly remains an
important direction for future work.

Within this framework, we first showed that the Adam--Gibbs relation follows
naturally from the scarcity of such bottleneck configurations. The entropy
deficit of the neck directly determines the waiting time for a successful
transition, leading to the characteristic dependence of the relaxation time
on the configurational entropy. This provides a transparent geometric
interpretation of the Adam--Gibbs relation in terms of connectivity in
configuration space.

We then formulated the problem using the Mori--Zwanzig projection operator
formalism with two slow variables: a progress coordinate \(X\) describing
motion between basins, and a neck variable \(Q\) describing the instantaneous
connectivity of configuration space. Eliminating the neck variable yields a
closed generalized Langevin equation in which the slow component of the
memory kernel is governed by the dynamics of the entropic bottleneck. This
provides a direct dynamical route from the neck picture to activated
relaxation, embedding the Adam--Gibbs relation within a microscopic
memory-function framework. We note that the neck variables $Q$ is a collective variables
which would be connected to the fluctuating density field, and hence can be connected
to the mode coupling theory (MCT). However, we have made no attempt to quantify this
correlation which at present is not available.This is clearly a weakness of the theory.

A key result of the present work is the decomposition of the relaxation rate
into the product
\begin{equation}
k(T) \sim N_{\rm connections}(T)\, P_{\rm viable}(T),
\end{equation}
which separates the total number of available inter-basin pathways from the
probability that a given pathway is dynamically viable. This factorization
provides a natural explanation for the crossover from Rosenfeld scaling to
Adam--Gibbs behavior. At high temperatures, the abundance of connections,
controlled by the excess entropy, governs the dynamics, while at low
temperatures the rarity of viable cooperative pathways dominates, leading
to activated behavior controlled by the configurational entropy.

The present formulation also clarifies the role of fragility. As temperature
decreases, the progressive narrowing and reduced persistence of the
entropic neck lead to a rapid growth of the characteristic timescale
\(\tau_Q(T)\), and hence of the structural relaxation time. Fragile systems,
in which the configurational entropy decreases rapidly, exhibit a sharper
collapse of viable connectivity and a correspondingly stronger growth of
relaxation times, while strong systems retain broader and more persistent
necks over a wider temperature range.

Overall, the entropic-neck picture provides a unified perspective that
connects geometric constraints in configuration space, thermodynamic
quantities such as configurational entropy, and dynamical quantities such
as memory kernels and relaxation times. It establishes a possible link between
Zwanzig’s description of diffusion in constrained geometries, the
Adam--Gibbs relation, and modern theories of glassy dynamics, including
RFOT. The framework developed here suggests that the central bottleneck
in glassy relaxation is fundamentally entropic and dynamical in origin,
arising from the scarcity and temporal persistence of viable pathways in
configuration space.

It will also be interesting to relate the present framework to vapor-deposited
ultrastable glasses, where molecules can access deeper basins and where the
corresponding entropic-neck barriers may be altered by the enhanced packing
and anisotropy of the deposited structure \cite{Ediger1, Kushal1, Kushal2, WolynesUltraStable}

Future work may also extend this approach to compute explicit forms of the neck
memory kernel, incorporate spatial heterogeneity, and establish more
quantitative connections with simulations and experiments probing dynamic
facilitation, cooperativity, and relaxation spectra.

\section*{Supplementary Material}

The Supplementary Material (\textbf{appended below, after the bibliography}) presents a memory-function description of fictive temperature evolution during cooling. It includes the derivation of the enthalpy memory kernel, the connection between configurational relaxation and calorimetric response, and the relation between neck-controlled dynamics, frequency-dependent specific heat, and nonequilibrium glass formation. A mode-coupling theory expression is also provided here.




\clearpage

\section*{Supplementary Material}

\setcounter{section}{0}
\setcounter{equation}{0}
\setcounter{figure}{0}
\setcounter{table}{0}

\renewcommand{\thesection}{SI-\arabic{section}}
\renewcommand{\thesubsection}{SI-\arabic{section}.\arabic{subsection}}
\renewcommand{\theequation}{S\arabic{equation}}
\renewcommand{\thefigure}{S\arabic{figure}}
\renewcommand{\thetable}{S\arabic{table}}



\section{Fictive temperature and the enthalpy memory kernel}

In the main text we have focused on the equilibrium and near-equilibrium
crossover from Rosenfeld-type excess-entropy scaling to Adam--Gibbs-type
activated relaxation. The central physical object there is the fluctuating
entropic neck in configuration space. The purpose of this Supplementary
Material is more limited. We show how the same slow structural relaxation
may also be represented in a calorimetric language through the fictive
temperature and an enthalpy memory kernel. This extension is not required
for the derivation of the Rosenfeld--Adam--Gibbs crossover presented in the
main text, but it provides a useful connection between neck-controlled
relaxation, nonequilibrium cooling, and calorimetric response.

We begin with the Tool--Narayanaswamy description of the slowly relaxing
configurational enthalpy. The total enthalpy is separated into a rapidly
equilibrated vibrational part and a slowly relaxing configurational part,
\begin{equation}
H(t)=H_{\rm vib}[T(t)]+H_{\rm conf}(t).
\label{eq:S_H_decomp}
\end{equation}
The vibrational degrees of freedom are assumed to adjust rapidly to the bath
temperature \(T(t)\), whereas the configurational contribution relaxes on
the structural time scale. The fictive temperature \(T_f(t)\) is introduced
by writing the configurational enthalpy as the equilibrium configurational
enthalpy of a liquid at temperature \(T_f(t)\),
\begin{equation}
H_{\rm conf}(t)=H_{\rm conf}^{\rm eq}[T_f(t)] .
\label{eq:S_fictive_def}
\end{equation}
For small departures from equilibrium, the excess configurational enthalpy
relative to the instantaneous equilibrium value at the bath temperature is
therefore
\begin{equation}
\Delta H(t)
\equiv
H_{\rm conf}(t)-H_{\rm conf}^{\rm eq}[T(t)]
\simeq
\Delta C_p \,[T_f(t)-T(t)] ,
\label{eq:S_DeltaH}
\end{equation}
where \(\Delta C_p\) is the configurational contribution to the heat capacity
at constant pressure. Thus \(T_f(t)\) measures the structural state of the
liquid and its lag behind the imposed thermal history.

A general memory-function form of the configurational enthalpy relaxation may
be written as
\begin{equation}
\frac{dH_{\rm conf}(t)}{dt}
=
-\int_0^t dt'\,
K_H(t-t';T_f)
\left[
H_{\rm conf}(t')-H_{\rm conf}^{\rm eq}(T(t'))
\right] .
\label{eq:S_H_memory}
\end{equation}
Here \(K_H(t;T_f)\) is the enthalpy memory kernel associated with the
structural state characterized by \(T_f\). The dependence of \(K_H\) on
\(T_f\) is important because, during nonequilibrium cooling, the structural
relaxation is governed not only by the instantaneous bath temperature but
also by the history-dependent configuration of the liquid.

Using Eq.~(\ref{eq:S_DeltaH}), Eq.~(\ref{eq:S_H_memory}) gives the
corresponding memory equation for the fictive temperature,
\begin{equation}
\frac{dT_f(t)}{dt}
=
-\int_0^t dt'\,
K_H(t-t';T_f)
\left[
T_f(t')-T(t')
\right] .
\label{eq:S_Tf_memory}
\end{equation}
This is the memory-function form of fictive-temperature dynamics. It
generalizes the usual single-relaxation-time equation by allowing the
structural response to depend on the full thermal history.

If the enthalpy kernel is short-ranged in time, it may be replaced by a
Markovian kernel,
\begin{equation}
K_H(t-t';T_f)
\simeq
\frac{1}{\tau[T(t),T_f(t)]}\,\delta(t-t') ,
\label{eq:S_Markov_kernel}
\end{equation}
and Eq.~(\ref{eq:S_Tf_memory}) reduces to the familiar
Tool--Narayanaswamy form
\begin{equation}
\frac{dT_f(t)}{dt}
=
-\frac{T_f(t)-T(t)}
{\tau[T(t),T_f(t)]} .
\label{eq:S_TN_equation}
\end{equation}
Thus the standard fictive-temperature equation may be viewed as a
single-mode or Markovian approximation to the more general
enthalpy-memory equation.

\section{Relation to enthalpy fluctuations}

The enthalpy--enthalpy time correlation function is the central equilibrium
quantity underlying this description. We define the slow configurational
enthalpy correlation function by
\begin{equation}
C_{HH}(t;T_f)
=
\left\langle
\delta H_{\rm conf}(t)\,\delta H_{\rm conf}(0)
\right\rangle_{T_f},
\label{eq:S_CHH_def}
\end{equation}
where
\begin{equation}
\delta H_{\rm conf}(t)
=
H_{\rm conf}(t)
-
\left\langle H_{\rm conf}\right\rangle_{T_f}.
\label{eq:S_deltaH_def}
\end{equation}
The average is taken over an equilibrium ensemble corresponding to the
structural state specified by \(T_f\). Within linear response, the slow part
of the enthalpy relaxation kernel is determined by the decay of this
correlation function. With a convenient normalization one may write
\begin{equation}
K_H(t;T_f)
\simeq
-
\frac{1}{C_{HH}(0;T_f)}
\frac{d C_{HH}(t;T_f)}{dt}.
\label{eq:S_kernel_CHH_norm}
\end{equation}
Equivalently, since
\begin{equation}
C_{HH}(0;T_f)
=
k_B T_f^2 \Delta C_p ,
\label{eq:S_CHH_static}
\end{equation}
one may write
\begin{equation}
K_H(t;T_f)
\simeq
-
\frac{1}{k_B T_f^2 \Delta C_p}
\frac{d C_{HH}(t;T_f)}{dt}.
\label{eq:S_kernel_CHH}
\end{equation}
The precise numerical prefactor depends on the convention used to define the
memory kernel, but the physical content is independent of this convention:
the fictive-temperature dynamics and the frequency-dependent heat capacity
are controlled by the same slow configurational enthalpy fluctuations.

This also clarifies the relation to the main text. In the entropic-neck
description, structural relaxation is slowed by the rarity and persistence of
viable configurational bottlenecks. In the calorimetric description, the same
slow structural relaxation appears as a long-lived enthalpy memory kernel.
The two descriptions emphasize different observables, but they refer to the
same underlying configurational dynamics.

\section{Mode-coupling representation of the enthalpy kernel}

A schematic microscopic representation of the slow configurational enthalpy
can be obtained by expressing it in terms of density fluctuations. For a
liquid with pair interaction \(v(r)\), the potential part of the enthalpy may
be written, up to one-body and constant terms, as
\begin{equation}
\delta H_{\rm conf}(t)
\simeq
\frac{1}{2V}
\sum_{\mathbf{k}}
v_k\,
\delta \rho_{\mathbf{k}}(t)
\delta \rho_{-\mathbf{k}}(t),
\label{eq:S_H_density}
\end{equation}
where \(v_k\) is the Fourier transform of the interaction potential, \(V\) is
the volume, and \(\delta \rho_{\mathbf{k}}(t)\) is the Fourier component of
the density fluctuation.

The normalized intermediate scattering function is defined as
\begin{equation}
\phi_k(t)
=
\frac{F(k,t)}{S(k)}
=
\frac{
\left\langle
\delta \rho_{\mathbf{k}}(t)
\delta \rho_{-\mathbf{k}}(0)
\right\rangle}
{S(k)} ,
\label{eq:S_phi_def}
\end{equation}
where \(S(k)=F(k,0)\) is the static structure factor. Using a standard
Gaussian or mode-coupling factorization of four-point density correlations,
one obtains the schematic expression
\begin{equation}
C_{HH}(t;T_f)
\simeq
\frac{1}{2V}
\sum_{\mathbf{k}}
v_k^2 S^2(k)\phi_k^2(t).
\label{eq:S_CHH_MCT}
\end{equation}
Equation~(\ref{eq:S_CHH_MCT}) shows that the enthalpy correlation function
inherits its slow decay from the same density modes that control the
mode-coupling description of structural relaxation.

Substitution of Eq.~(\ref{eq:S_CHH_MCT}) into
Eq.~(\ref{eq:S_kernel_CHH}) gives the corresponding schematic
mode-coupling representation of the enthalpy memory kernel,
\begin{equation}
K_H(t;T_f)
\simeq
-
\frac{1}{2V k_B T_f^2 \Delta C_p}
\frac{d}{dt}
\sum_{\mathbf{k}}
v_k^2 S^2(k)\phi_k^2(t).
\label{eq:S_K_MCT}
\end{equation}
This expression should not be regarded as a closed microscopic theory by
itself. Its purpose is to show that the calorimetric memory kernel is
governed by the same slow density correlations that enter the usual
mode-coupling treatment.

\section{Breakaway of the fictive temperature}

The fictive temperature follows the bath temperature as long as the
structural relaxation time is short compared with the externally imposed
cooling time scale. For a linear cooling protocol,
\begin{equation}
T(t)=T_0-qt ,
\qquad q>0 ,
\label{eq:S_cooling}
\end{equation}
the system begins to fall out of equilibrium when the intrinsic structural
relaxation time becomes comparable to the time scale over which the bath
temperature changes appreciably.  Thus the relevant time scale is a
temperature scale divided by the cooling rate, not \(1/q\).

A simple operational criterion is
\begin{equation}
\tau_\alpha(T^{*})
\sim
\frac{T^{*}}{q},
\label{eq:S_breakaway}
\end{equation}
where \(T^{*}\) denotes the breakaway temperature at which \(T_f(t)\) first
departs measurably from \(T(t)\). More refined definitions replace
\(T^{*}/q\) by the local time scale associated with the imposed cooling
protocol, but Eq.~(\ref{eq:S_breakaway}) is sufficient for the present
qualitative discussion.

Equation~(\ref{eq:S_K_MCT}) shows that the enthalpy kernel becomes
long-lived when the density correlators \(\phi_k(t)\) develop slow
relaxation. In the idealized mode-coupling description this slowing down is
associated with the approach to the mode-coupling crossover temperature
\(T_c\). In real liquids the ideal transition is avoided by activated
processes, but the same temperature window often marks a pronounced change
in the character of relaxation. Thus the calorimetric breakaway scale
\(T^{*}\) may lie in the same broad crossover regime as \(T_c\), especially
in fragile glass formers where the relaxation time grows rapidly with
decreasing temperature.

This statement should not be interpreted as an identity between \(T^{*}\)
and \(T_c\). The breakaway temperature depends on the cooling rate, whereas
\(T_c\) is an equilibrium crossover scale extracted from dynamical
correlations or from mode-coupling fits. The point is instead that the
enthalpy memory kernel provides a formal route for connecting nonequilibrium
calorimetric lag to the slowing down of microscopic density modes.

\section{Relation to the entropic-neck picture}

The enthalpy-memory formulation and the entropic-neck formulation are
complementary. The main text describes structural relaxation as motion
through rare, fluctuating bottlenecks in configuration space. Elimination of
the neck variable produces a slow memory kernel for the progress coordinate.
The present Supplement describes the same slow configurational relaxation
from the viewpoint of enthalpy fluctuations and fictive-temperature
dynamics.

In the high-temperature regime, density correlations decay rapidly, the
enthalpy kernel is short-ranged in time, and \(T_f(t)\) follows \(T(t)\)
closely. In the deeply supercooled regime, density correlations and neck
variables become long-lived, the enthalpy kernel develops a slow tail, and
the fictive temperature falls out of equilibrium. Thus the calorimetric
memory kernel gives an experimentally accessible representation of the same
slow configurational dynamics that, in the main text, appears as
neck-controlled memory and Adam--Gibbs-type activated relaxation.

\end{document}